\documentclass{aa}  

\usepackage{graphicx,xcolor}
\usepackage[normalem]{ulem}
\usepackage{physics}
\usepackage{txfonts}
\usepackage{float}

\begin{document}

   \title{3D simulations of AGB stellar winds — II. Ray-tracer implementation and impact of radiation on the outflow morphology}
   \titlerunning{Approximate radiative transfer techniques solving for the radiative acceleration}
   \author{
          M. Esseldeurs\inst{1} \and
          L. Siess\inst{2} \and
          F. De Ceuster\inst{1} \and
          W. Homan\inst{2} \and
          J. Malfait\inst{1} \and\\
          S. Maes\inst{1} \and
          T. Konings\inst{1} \and
          T. Ceulemans\inst{1} \and
          L. Decin\inst{1, 3}
          }

   \institute{Instituut voor Sterrenkunde, KU Leuven, Celestijnenlaan 200D, 3001 Leuven, Belgium\\
              \email{mats.esseldeurs@kuleuven.be}
         \and
             Institut d’Astronomie et d’Astrophysique, Université Libre de Bruxelles (ULB), CP 226, 1050 Brussels, Belgium
         \and
             School of Chemistry, University of Leeds, Leeds LS2 9JT, UK
             }

   \date{Received 1 March 2023 / Accepted 17 April 2023}
 
   \abstract
   {Stars with an initial mass below $\sim8\,{\rm M}_\odot$ evolve through the asymptotic giant branch (AGB) phase, during which they develop a strong stellar wind, due to radiation pressure on newly formed dust grains. Recent observations have revealed significant morphological complexities in AGB outflows, which are most probably caused by the interaction with a companion.}
   {We aim for a more accurate description of AGB wind morphologies by accounting for both the radiation force in dust-driven winds and the impact of a companion on the AGB wind morphology.}
   {We present the implementation of a ray tracer for radiative transfer in the smoothed particle hydrodynamics (SPH) code \textsc{Phantom}. Our method allows for the creation of a 3D map of the optical depth around the AGB star. The effects of four different descriptions of radiative transfer, with different degrees of complexity, are compared:  the free-wind approximation, the geometrical approximation, the Lucy approximation, and the attenuation approximation. Finally, we compare the Lucy and attenuation approximation to predictions with the 3D radiative transfer code \textsc{Magritte}.}
   {The effects of the different radiative transfer treatments are analysed considering both a low and high mass-loss rate regime, and this both in the case of a single AGB star, as well as for an AGB binary system. For both low and high mass-loss rates, the velocity profile of the outflow is modified when going from the free-wind to the geometrical approximation, also resulting in a different wind morphology for AGB binary systems. In the case of a low mass-loss rate, the effect of the Lucy and attenuation approximation is negligible due to the low densities but morphological differences appear in the high mass-loss rate regime. 
   By comparing the radiative equilibrium temperature and radiation force to the predictions from \textsc{Magritte}, we show that for most of the models, the Lucy approximation works best. Although, close to the companion, artificial heating occurs and it fails to simulate the shadow cast by the companion.
   The attenuation approximation leads to stronger absorption of the radiation field, yielding a lower equilibrium temperature and weaker radiation force, but it produces the shadow cast by the companion.
   From the predictions of the 3D radiative transfer code \textsc{Magritte}, we also conclude that a radially directed radiation force is a reasonable assumption.}
   {The radiation force plays a critical role in dust-driven AGB winds, impacting the velocity profile and morphological structures. For low mass-loss rates, the geometrical approximation suffices, however for high mass-loss rates, a more rigorous method is required. Among the studied approaches, the Lucy approximation provides the most accurate results, although it does not account for all effects.}


   \maketitle

\section{Introduction}\label{sec:introduction}
    Asymptotic giant branch (AGB) stars are the late evolutionary stage of low- and intermediate-mass stars (0.8 M$_\odot \lesssim$ M$_\star$ $\lesssim$ 8 M$_\odot$). These stars exhibit mass-loss rates that range from $10^{-8}$ up to $10^{-5} \,{\rm M_\odot \, yr}^{-1}$, with terminal wind speeds of $5-30 \,{\rm km \, s}^{-1}$ \citep{Habing2004, Ramstedt2008}. To achieve such outflows, a mechanism is needed to overcome the stellar gravitational attraction. For AGB stars, the wind is believed to be a pulsation-enhanced dust-driven wind. A complex interplay between strong convection in the AGB atmosphere, and large-amplitude long-period pulsations, forms shock waves in the atmosphere \citep{freytag2008, freytag2017, freytag2023}. These shocks levitate the gas into sufficiently cool regions where it is able to condensate into dust. Dust particles can efficiently absorb stellar radiation, such that they are pushed outwards by the radiation force. When moving outwards, the dust collides with the surrounding gas and drags it along. This creates an efficient mechanism for mass loss around AGB stars \citep{Lamers1999, Hofner2018}.
    
    High-resolution observations of AGB stars have revealed complex structures in their outflows \citep{Ramstedt2014, Kervella2016, Decin2020}. One of the leading hypotheses to explain these morphologies is the presence of a binary companion, gravitationally shaping the outflow into complex morphologies. To investigate this hypothesis, 3D hydrodynamic studies have been performed using both grid-based and smoothed particle hydrodynamics (SPH) codes \citep[e.g.][]{Theuns1993, Mastrodemos1999, Kim2012, Saladino2018, Saladino2019, Maes2021, Malfait2021, Aydi2022, Lee2022}.
    Most of these studies use the so-called free-wind approximation where the gravity of the mass-losing star is ignored, and none of the complexities of the wind-launching mechanism are included.
    To improve the modelling of the outflow simulations, recent attempts have been made to include more of the underlying physical mechanisms.
    Using a grid-based code, \cite{Chen2017, Chen2020} considered pulsations and an approximate form of dust opacity and radiative transfer, showing that these implementations alter the resulting outflow morphologies and could lead to the formation of circumbinary disks. In SPH codes, \cite{Aydi2022} simulated pulsations to launch the wind following the 1D modelling of \cite{bowen1988}, while \cite{Siess2022} implemented dust nucleation.
    
    Previous studies \citep[e.g.][]{Theuns1993,Mastrodemos1999,Kim2012,Maes2021,Malfait2021} showed that the 
    the structures, emerging from the interaction of a companion with the AGB wind, depend on the relative magnitude of the wind and orbital velocities.
    Understanding structure formation with a mass-losing AGB star is therefore tightly linked with understanding the velocity of the wind.
    To model this wind velocity realistically, an accurate acceleration prescription is required, including a realistic description of the radiation force.
  
    Radiative transfer is thus a key ingredient for stellar wind models, as it provides the driving force in dust-driven winds. However, except for 'star-in-the-box' type simulations \citep{freytag2023}, performing full radiative transfer calculations on-the-fly in spatially extended simulations is not yet feasible due to the large computational cost. Therefore, one needs to resort to approximate radiative transfer descriptions to investigate the effect of radiation on the shaping of the wind on large scales, potentially perturbed by a companion.
    
    In this paper, we present the implementation of a ray tracer in the SPH code \textsc{Phantom}, and use it to define four different radiative transfer approximations, each of which yielding a different prescription for the radiation force. Each approximation increases the complexity, from no radiative transfer (free-wind approximation), to geometrical dilution (geometrical approximation), to accounting for radiation not exclusively coming from the AGB star (Lucy approximation), and attenuation of the stellar radiation (attenuation approximation). We analyse the effects of the approximations on the velocity profiles and morphological structures, considering two mass-loss rate regimes, both in the case of a single AGB star, as well as in an AGB binary system. Finally, we compare the Lucy and attenuation approximations to the results of the 3D radiative transfer code \textsc{Magritte}.
    
    The outline of this paper is as follows. In Sect.~\ref{sec:modelsetup}, we present the four different radiative transfer prescriptions investigated in this study, as well as their numerical implementation. In Sect.~\ref{sec:single}, the effects of these prescriptions on the velocity profile of a single AGB star are investigated. In Sect.~\ref{sec:binary}, the changes on the morphological structures of a binary system with a primary mass-losing AGB star are analysed. A comparison between the approximations and the full 3D radiative transfer code \textsc{Magritte} is given in Sect.~\ref{sec:discussion}, and the main results are summarized in Sect.~\ref{sec:summary}.

\section{Model and setup}\label{sec:modelsetup}

    \subsection{Smoothed particle hydrodynamics}\label{sec:SPH}
        The smoothed particle hydrodynamics (SPH) code \textsc{Phantom} \citep{Price2010, Lodato2010, Price2018} is used for our 3D hydrodynamic simulations.
        In the framework of SPH \citep{Gingold1977,Lucy1977}, the density distribution of a particle, the equations of motion, and energy conservation read \citep{Price2018}:
        \begin{equation}
            \ \ \ \rho_i=\sum_j m_j W\left(\left|\boldsymbol{r}_i-\boldsymbol{r}_j\right|, h_i\right) \label{eq:consofmass} \ ,
        \end{equation}\vspace*{-1.5em}
        \begin{equation}
            \frac{\mathrm{d} \boldsymbol{v}_i}{\mathrm{~d} t} =
                -\sum_j m_j\Bigg[\frac{P_i+q_{i}}{\rho_i^2 \Omega_i} \nabla_i W_{i j}\left(h_i\right)+\frac{P_j+q_j}{\rho_j^2 \Omega_j} \nabla_i W_{ij}(h_j)\Bigg]+\boldsymbol{a}_{\text {ext}, i}\label{eq:consofmom} \ ,
        \end{equation}\vspace*{-2.5em}
        \begin{equation}
            \frac{\mathrm{d} u_i}{\mathrm{~d} t} = \frac{P_i}{\rho_a^2 \Omega_i} \sum_j m_j (\boldsymbol{v}_{i}-\boldsymbol{v}_{j}) \cdot \nabla_a W_{a b}\left(h_a\right)+\Lambda_{\mathrm{shock}}-\frac{\Lambda_{\mathrm{cool}}}{\rho}\label{eq:consofener} \ ,
        \end{equation}
        where $i$ denotes the SPH particle, and $j$ its neighbouring particles. Particles are defined by their (constant) mass $m$, position  $\boldsymbol{r}$, velocity $\boldsymbol{v}$, and specific internal energy $u$. The calculation of the local density $\rho$ and pressure $P$ requires knowledge of the neighbours, which are found within the support of the smoothing kernel $W$ (particles within $R_{\rm kern}h$, where $R_{\rm kern}$ is the (dimensionless) kernel radius and $h$ the smoothing length). We use the $M_4$ cubic spline kernel, which vanishes outside a radius $r = 2 h$. This leads on average to a number of neighbours equal to $N_{\rm neigh} = 57.9$ \citep{Price2018}. In Eq.~\eqref{eq:consofmom}, $\boldsymbol{a}_{\mathrm{ext},i}$ represents the external accelerations applied to the particle $i$. In this setup, the forces are the gravitational attraction of the star(s) and of the additional radiation force on the particle. In a binary system, it is given by 
        \begin{equation}\label{eq:acceleration}
            \boldsymbol{a}_{\text {ext},i} = -\frac{GM_\mathrm{AGB}}{r_{i,1}^2}\left(1-\Gamma_i\right)\hat r_{i,1} - \frac{GM_\mathrm{comp}}{r_{i,2}^2}\hat r_{i,2}\ ,
        \end{equation}
        where $\boldsymbol{r}_{i,1}$ and $\boldsymbol{r}_{i,2}$ are the distances from the position of the $i$'th particle to the AGB star and the companion, respectively. $\Gamma_i$ is the Eddington factor, which is given by
        \begin{equation}\label{eq:Gamma}
            \Gamma_i = \frac{(\kappa_d + \kappa_g)F_i \, r_{i,1}^2}{G M_\mathrm{AGB}\, c}\ ,
        \end{equation}
        in which $\kappa_d$ and $\kappa_g$ are the opacity of the dust and gas, respectively, and $F_i$ is the flux coming from the AGB star, reaching the $i$'th particle. The widely adopted free-wind approximation implies setting $\Gamma_i$ equal to one, not taking into account the potentially very complex behaviour of $\Gamma$ (see Sect.~\ref{sec:FreeWind} for a more detailed description).
        
        The calculation of the dust opacity is complex. Several prescriptions are available in \textsc{Phantom}, including the complex nucleation theory \citep{Siess2022}, which is based on the theory of moments developed by \cite{Gail2013}. However, for the purpose of this paper, to reduce the computational cost and the complexity of the interpretation of the results, we use the simplified analytic dust opacity given by \cite{bowen1988}
        \begin{equation}\label{eq:kappadust}
            \kappa_d(T_\mathrm{eq}) = \frac{\kappa_\mathrm{max}}{1+\exp[(T_\mathrm{eq}-T_\mathrm{cond})/\delta]} \ ,
        \end{equation}
        where $T_\mathrm{eq}$ is the dust temperature, $T_{\rm{cond}} = 1\,200$~K the dust condensation temperature, $\kappa_\mathrm{max} = 6\,{\rm cm}^{2}\,{\rm g}^{-1}$ the maximal dust opacity, and $\delta = 60$~K the temperature range over which dust condensation occurs. These parameters can be changed to model different types of dust, where this setup describes carbon-rich dust \citep{bowen1988}.
        For the gas opacity, a constant value of $\kappa_{g} = 2\times10^{-4}\,{\rm cm}^{2}\,{\rm g}^{-1}$ is chosen \citep{bowen1988}.

         In Eq.~(\ref{eq:consofener}), $\Lambda_\mathrm{shock}$ represents the energy dissipation rate required to give the correct entropy increase in shocks. It consists of viscous shock heating, artificial thermal conductivity, and artificial resistivity if magnetic fields are included \citep[for details, see][]{Price2018}.
         The cooling rate is represented by $\Lambda_\mathrm{cool}$. Two processes have been included in our simulations. They correspond to the \cite{bowen1988} prescription which is expressed as
        \begin{equation}\label{eq:bowen_cooling}
            \Lambda_\mathrm{Bowen} = \frac{3R}{2\mu}\frac{T_{g}-T_\mathrm{eq}}{C'} \ ,
        \end{equation}
        where $R$ is the gas constant, $\mu$ the mean molecular weight, and $C'$ the parametric cooling rate, taken to be $3 \times 10^{-5} \,{\rm g \, s \,cm}^{-3}$ \citep{bowen1988}. This expression, which is proportional to the temperature difference between the gas and the dust, was constructed to mimic diffusion between the two species. Furthermore, cooling by neutral hydrogen, $\Lambda_\mathrm{H}$, is also considered, following the formula given by \cite{Spitzer1978}. The total cooling rate is thus given by $\Lambda_\mathrm{cool} = \Lambda_\mathrm{Bowen} + \Lambda_\mathrm{H}$.

        \subsection{Radiative transfer approximations}\label{sec:RTapprox}
            The transport of energy via radiation can be considered along specific directions, often referred to as rays. The frequency-dependent (indicated with the subscript $\nu$) radiative transfer equation along a ray reads
        	\begin{equation}
        	\frac{\dd I_\nu}{\dd s} = \eta_\nu - \alpha_\nu I_\nu\ .
        	\label{eq:RT}
        	\end{equation}
        	Here, $I_\nu$ represents the intensity (i.e. the observable quantity), $\dd s$ is a path element along the ray, $\eta_\nu$ is the emission coefficient, which is a measure for the radiative energy that is gained along to the ray, and $\alpha_\nu$ is the absorption coefficient, which quantifies the radiative energy lost along the ray ($\alpha_\nu=\kappa_\nu \rho$). By defining the frequency-dependent optical depth, $\tau_\nu$, as 
        	\begin{equation}
        	\dd \tau_\nu = \alpha_\nu \dd s \ ,
        	\end{equation}
        	the transfer equation can be re-written as 
        	\begin{equation}
        	\frac{\dd I_\nu}{\dd \tau_\nu} = S_\nu - I_\nu\ ,
        	\end{equation}
        	where $S_\nu = \eta_\nu/\alpha_\nu$ is referred to as the source function.
            The transfer equation can formally be solved, yielding
        	\begin{eqnarray}
        	I_\nu(s) = I_\nu(0)\, e^{-\tau_\nu(s)}+\int_{0}^{\tau_\nu(s)}S_\nu\, e^{-(\tau_\nu(s)-\tau)} \dd\tau\ .
        	\label{eq:RTformal}
        	\end{eqnarray}
        	For a homogeneous medium, the source function does not change throughout the medium, and can be moved outside of the integral, yielding the solution
        	\begin{equation}
        	I_\nu = I_\nu(0) e^{-\tau_\nu}+S_\nu (1-e^{-\tau_\nu})\ .
        	\label{eq:RTgrey}
        	\end{equation}
        	From the intensity, we define the mean intensity $J_\nu$ and flux $F_\nu$
        	\begin{align}
        	    J_\nu  & = \frac{1}{4\pi} \oint I_\nu(\theta,\phi) \ \dd\Omega\ , \label{eq:Imean} \\
        	    F_\nu & = \oint I_\nu(\theta,\phi) \cos \theta\ \dd\Omega \ , \label{eq:Flux}
        	\end{align}
            where $\dd\Omega = \sin\theta \dd\theta \dd\phi$ is the differential solid angle, and $\theta$ the angle between the direction of the intensity and a normal vector of the surface through which we are considering the flux.
        	Under the condition of local thermodynamic equilibrium (LTE), the source function equals the Planck function, $S_\nu = B_\nu(T)$, and   
        	\begin{equation}
        	    \int_0^\infty \kappa_\nu B_\nu \, \dd\nu = \int_0^\infty \kappa_\nu J_\nu\, \dd\nu\ .
        	    \label{eq:rad_eq}
        	\end{equation}
        	In general, knowledge of the mean intensity, $J_\nu$, requires to solve the radiative transfer equations throughout the entire medium. However, some approximations can allow us to avoid this costly computation. 
        	In our study, we restrict ourselves to the grey case (i.e. ignoring any frequency dependence). Assuming radiative equilibrium, the frequency-integrated mean intensity and flux are given by 
        	\begin{eqnarray}
        	    J & = & \int_0^\infty J_\nu\, \dd\nu = \int_0^\infty B_\nu\, \dd\nu = \frac{\sigma_\mathrm{sb}}{\pi} \, T^4
        	    \label{eq:Jdef} \ , \\
        	    F & = & \int_0^\infty F_\nu\, \dd\nu \ ,
        	\end{eqnarray} 
            where $\sigma_\mathrm{sb}$ is the Stefan-Boltzmann constant. 
            The computation of the variables $F$ and $T_{\rm{eq}}$, using a full radiative transfer description, is too computationally demanding. In order to alleviate this problem, we investigate four different approximate descriptions, and evaluate their applicability for 3D stellar wind models.
            
        \subsubsection{Free-wind approximation}\label{sec:FreeWind}
            The free-wind approximation is the most drastic approximation, since no explicit treatment of the radiation transport is included, and $\Gamma$ (Eq.~\ref{eq:Gamma}) is simply set equal to one, implying the SPH particles do not feel the gravitational pull of the mass-losing star, as it is artificially balanced by the radiation force. Despite the crudeness of this approximation, it provides a simple way to launch a wind without requiring a full treatment of pulsations and dust formation. This approximation was introduced by \cite{Theuns1993} and has been widely adopted since, because of its simplicity \citep[e.g.][]{Mastrodemos1999,Kim2012,Liu2017,Saladino2019,Maes2021,Malfait2021,Lee2022}.

        \subsubsection{Geometrical approximation}\label{sec:Geometrical}
            Assuming a spherical star of radius $R_\star$, an SPH particle located at a distance $r$ from the source `sees' the star over an opening angle $\theta_M$, given by $\sin \theta_M = R_\star/r$.
            Furthermore, assuming that the star is isotropically emitting a constant intensity $I_\nu$, such that $I_\nu = J_\nu$, the flux received at the particle's location is 
            \begin{eqnarray}
                F_\nu(r) & =&  I_\nu\int_0^{2\pi} \dd\phi \int_0^{\theta_M} \cos\theta \sin\theta \, \dd\theta \nonumber \\
                & = & \pi I_\nu \sin^{2} \theta_M \ = \ F_\nu(R_\star) \, \frac{R_\star^2}{r^2}\ .
            \end{eqnarray}
            In the grey approximation, neglecting all frequency dependencies, the expression for $\Gamma$ (Eq.~\ref{eq:Gamma}) takes the standard form 
            \begin{equation}\label{eq:GammaGeometrical}
                \Gamma = \frac{(\kappa_d + \kappa_g) L_\mathrm{AGB}}{4\pi c G M_\mathrm{AGB}}\ ,
            \end{equation}
            where $L_\mathrm{AGB} = 4\pi R_\star^2 F(R_\star)$ is the AGB luminosity. In this expression, the Eddington factor depends on the dust opacity, which requires the dust temperature. From Eq.~(\ref{eq:Imean}), the mean intensity reads
            \begin{eqnarray}
                J_\nu(r) & = & \frac{I_\nu}{4\pi}\int_0^{2\pi} \dd\phi \int_0^{\theta_M} \sin \theta \, \dd\theta = \frac{1}{2} I_\nu\, (1-\cos\theta_M) \nonumber \\
                & = & \frac{1}{2}\left[1-\sqrt{1-\left(\frac{R_\star}{r}\right)^2}\right] I_\nu = W(r)\ I_\nu \ ,
            \end{eqnarray}
            where $W(r)$ is often referred to as the geometrical dilution factor.
            Assuming radiative equilibrium, the local mean intensity can be related to the radiative equilibrium temperature by
            \begin{equation}\label{eq:TdustGeometrical}
                T_\mathrm{eq}^4(r) \ = \ \frac{\pi}{\sigma_\mathrm{sb}} J(r) \ = \ \frac{1}{2}\left(1-\sqrt{1-\left(\frac{R_\star}{r}\right)^2}\right) T_\star^4\ .
            \end{equation}
            This temperature is associated to the dust, because it absorbs most of the radiation due to its high intrinsic opacity. In this simple approximation, $\Gamma$ and $T_\mathrm{eq}$ can both be evaluated locally, making it easy to implement in a hydrodynamics code. However, we should emphasize that these expressions heavily rely on the assumptions of spherical symmetry, radiative equilibrium, and that the local mean intensity is dominated by the intensity of the AGB star. This approach was used, for instance, by \cite{Aydi2022}.

        \subsubsection{Lucy approximation}\label{sec:Lucy}
            Assuming spherical symmetry, local thermodynamic equilibrium \citep[LTE; which implicitly assumes that collisions dominate over radiative processes, see e.g.][Sects.~8.1 and 8.2]{Gail2013}, and an optically thin extended envelope, an improved prescription for $J$ can be obtained, which lifts the assumption that $J$ is dominated by the intensity of the AGB star. Following \cite{lucy1971, Lucy1976} (also described in Appendix~A1.2 of \citealp{Gail2013}), the frequency-integrated mean intensity $J$, as a function of distance $r$ from the central star, reads \citep[Eq.~12 from ][]{lucy1971}
            \begin{equation}
                J(r) = \left[\frac{1}{2}\left(1-\sqrt{1-\left(\frac{R_\star}{r}\right)^2}\right) + \frac{3}{4} \tau_L\right]J(R_\star) \ ,
                \label{eq:J_lucy}
            \end{equation}
            where the Lucy optical depth $\tau_L$ is given by 
            \begin{equation}\label{eq:tauL}
                \tau_L = \int_r^\infty (\kappa_d+\kappa_g) \, \rho \left(\frac{R_\star}{r'}\right)^2 \dd r'\ .
            \end{equation}
            The Lucy optical depth can physically be interpreted as quantifying the radiation that is absorbed by the surrounding envelope, outside a given distance $r$ from the AGB star.
            This radiation is re-emitted isotropically (because of radiative equilibrium) and provides a positive feedback contribution to the mean intensity at the considered location $r$.
            From Eq.~(\ref{eq:Jdef}), we immediately get the equilibrium temperature \citep[Eq.~3 of ][]{Lucy1976}:
            \begin{equation}\label{eq:TeqLucySpherical}
                T_\mathrm{eq}^4(r) \ = \ \frac{\pi}{\sigma_\mathrm{sb}} J(r) \ = \ \left[\frac{1}{2}\left(1-\sqrt{1-\left(\frac{R_\star}{r}\right)^2}\right) + \frac{3}{4} \tau_L\right] T_\star^4\ .
            \end{equation}
            If the spherical symmetry is broken, it is still possible to estimate the dust temperature at a given position by replacing the angle-independent radial coordinate ($r$) by  ($r, \theta, \phi$) in the calculation of $\tau_L$. In this generalized form, Eq.~\eqref{eq:TeqLucySpherical} writes
            \begin{equation}\label{eq:TeqLucy}
                T_\mathrm{eq}^4(r, \theta, \phi) = \left[\frac{1}{2}\left(1-\sqrt{1-\left(\frac{R_\star}{r}\right)^2}\right) + \frac{3}{4} \tau_L(r, \theta, \phi)\right] T_\star^4\ ,
            \end{equation}
            where $\theta$ and $\phi$ are the azimuthal and polar angles, respectively, indicating the direction of the ray, originating from the AGB star. This approach can better account for local dusty regions in the simulation.
            The Lucy approximation is often used in wind simulations to estimate the dust temperature \cite[e.g.][]{bowen1988,Saladino2018,Saladino2019,Lee2022}.

        \subsubsection{Attenuation approximation}\label{sec:attenuation}
            If the medium between the AGB star and an SPH particle is opaque and only absorbs the stellar radiation without radiating itself (i.e.\ $S_\nu = 0 $), then Eq.~(\ref{eq:RTgrey}) simplifies to
            \begin{equation}
                I_\nu = I_\nu(0)\, e^{-\tau_\nu}\ ,
            \end{equation}
            which does not depend on the symmetry of the problem. In the grey approximation, the flux at the particle's location then becomes 
            \begin{equation}
                F(r, \theta, \phi)
                \ = \  F(R_\star) \frac{R_\star^2}{r^2} \, e^{-\tau(r, \theta, \phi)}
                \ = \ \frac{L_\mathrm{AGB}}{4\pi r^2}\, e^{-\tau(r, \theta, \phi)} \label{eq:FAtten} \ ,
            \end{equation}
            where the optical depth $\tau$ is given by
            \begin{equation}\label{eq:tau}
                \tau(r, \theta, \phi) = \int_{R_\star}^r (\kappa_d+\kappa_g) \, \rho \, \dd r' \ .
            \end{equation}
            Likewise, the expression for the dust temperature now reads
            \begin{equation}
                T_\mathrm{eq}^4(r, \theta, \phi) = \frac{1}{2}\left(1-\sqrt{1-\left(\frac{R_\star}{r}\right)^2}\right) T_\star^4\, e^{-\tau(r, \theta, \phi)}  .\label{eq:TeqAtten}
            \end{equation}
            Here, the Lucy optical depth (Sect.~\ref{sec:Lucy}) has disappeared because there is no emission, and, hence, no external source of radiation can heat up the medium.
            This prescription was used, for instance, by \cite{Chen2017,Chen2020} in their study of the morphology of AGB outflows, perturbed by a stellar companion.
   
    \subsection{Ray-tracing implementation in \textsc{Phantom}}\label{sec:NumImp}
        The Lucy and attenuation approximations, described above, require the calculation of the optical depth $\tau$ or $\tau_L$. These are non-local quantities to be evaluated along the line of sight, connecting an SPH particle and the AGB star. For the computation of these quantities, we implemented the ray tracer described in this section\footnote{The routine can be found in the source code of \textsc{Phantom} at \url{https://github.com/danieljprice/phantom/blob/master/src/main/utils_raytracer.f90}}. More details, for instance, about the trade-offs made during development, can be found in Appendix~\ref{app:raytracer}.

        \begin{figure}
            \centering
            \includegraphics[width=\linewidth]{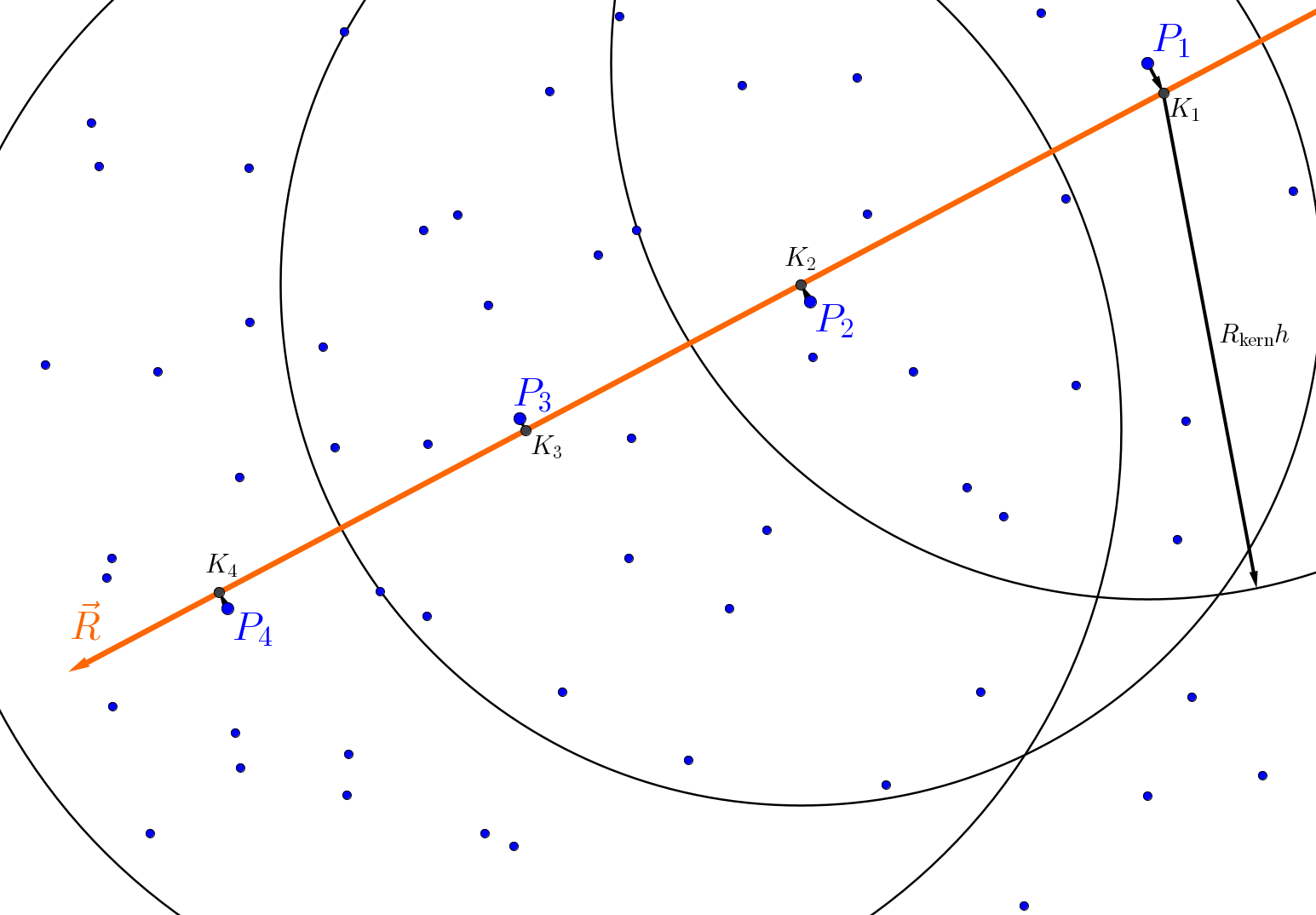}
            \caption{Visual representation of the ray-tracing algorithm. Starting from the upper right corner at point $P_1$ and following the direction of the ray, the subsequent point $P_2$ is determined within the sphere of influence of radius $R_\mathrm{kern}h$. The procedure is then repeated until the boundary of the domain is reached (see text for explanations).}
            \label{Fig:raytracing}
        \end{figure}
        
        \subsubsection{Algorithm}
        Since SPH particles are not located on a mesh, we opt for the meshless ray tracer that is implemented in \textsc{Magritte} \citep{DeCeuster2020b, DeCeuster2020a}. \textsc{Magritte} is a 3D radiative transfer code, developed to handle both mesh-based and meshless data, and can therefore work with SPH.
      
        The algorithm is visually explained in Fig.~\ref{Fig:raytracing}. Starting from a position $P_1$, a ray is traced in the $R$-direction. The algorithm searches for the nearest neighbours, using the information provided by the $k$d-tree of the SPH particles available in \textsc{Phantom}. From these neighbours, it selects the point closest to the ray, provided that the point lies in the direction of the ray and is within a distance $R_\mathrm{kern}h$, where $h$ is the smoothing length and $R_\mathrm{kern}=2$. This search is repeated until a given location, or the edge of the numerical domain, is reached.
        
        To calculate optical depths, a segment will be defined as the distance between two subsequent points projected on the ray, for instance, between $K_1$ and $K_2$ (the projections of $P_1$ and $P_2$, respectively) in Fig.~\ref{Fig:raytracing}. The optical depth can then be approximated by a sum over all segments created. using the ray-tracer.
        As such, the optical depth can be discretized as
        \begin{equation}
            \tau \ = \ \int_{R_\star}^r \dd \tau \ \approx \ \sum_{i} \Delta \tau_i \ = \ \sum_{i} \left(\frac{\kappa_i \,\rho_i + \kappa_{i+1} \,\rho_{i+1}}{2} \right) \, \Delta s_i \ ,
            \label{eq:tau_discretized}
        \end{equation}
        where the index, $i$, ranges over all points encountered along the ray, form the stellar surface at $R_\star$ to the radial distance from the AGB star, $r$, $\Delta s_i={\rm dist}({K}_{i+1},{K}_{i})$ is the length of the segment, $\kappa_i$, and $\rho_i$ are the opacity and density, respectively, at point $K_i$. $\Delta s_i$ can easily be determined from geometrical considerations, while $\kappa_i\,\rho_i$ requires a little more attention. It is evaluated using the SPH smoothing kernel, considering only the particles within two smoothing lengths, yielding
        \begin{equation}
            \kappa_i\,\rho_i = \sum_{j: \, |r_i-r_j| < 2 h_i} m_j\, \kappa_j W(r_i-r_j,h_j)\ .
            \label{eq:integrand}
        \end{equation}
        The opacity $\kappa_i\,\rho_i$ is computed at each point, $K_i$, along the ray and can then be used in Eq. \eqref{eq:tau_discretized} to give the integrated optical depth.

        \subsubsection{3D optical depth interpolation}

        To calculate the optical depth throughout the numerical domain, the most physically correct approach is to trace a ray from each SPH particle to the AGB star. This approach is similar to that of \cite{Kessel2000}. However, it is very computationally intensive and inefficient, because the optical depths obtained for the particles close to the AGB star are re-calculated whenever considering particles further out.
        One way around this, is to trace a predefined number of rays originating from the star, and interpolate the optical depth from the values of $\tau$ on these rays.
        The idea of tracing only a predefined number of rays, is common to most 3D ray-tracing radiative transfer algorithms \citep[see e.g.][]{Altay2013,DeCeuster2020b}.
        Starting from the surface of the star (at radius $R_\star$), rays are traced outwards, and the optical depth increments, $\Delta\tau_i$, are calculated, accumulated, and stored for each segment and for each ray.
        Now, to obtain the optical depth at each SPH particle, the following interpolation scheme is used.
        First, for each SPH particle, located at a distance $r$ from the star, the four rays passing closest to the particle are identified.
        Then, the optical depth along each of these four rays is linearly interpolated at the distance $r$. Finally, the optical depth values of the four rays (all evaluated at $r$) are used to estimate $\tau$ at the particle's location. The accuracy of this approach highly depends on the number of rays, as well as on their spatial distribution (see Appendix~\ref{app:rayinterpolation} for more details).
        
        To obtain a uniform distribution of rays in 3D, we use the \textsc{HEALPix} package \citep{Gorski2005}, specifically designed for this problem.
        \textsc{HEALPix} divides the 2-sphere into isolaterally distributed pixels of equal area. We use the unit vectors pointing to the centre of each pixel as the directions for the rays. In the nested scheme, \textsc{HEALPix} can refine the spatial division by increasing the number of pixels, such that each pixel is split into four. Each time this happens, the order of the scheme increases by one. So, at order zero, the 2-sphere is divided into 12 pixels, and for a general order $o$, the number of pixels is $n_\mathrm{rays} = 12 \times 4^o$. \textsc{HEALPix} also provides functions to obtain the pixel associated with every point on the 2-sphere. This means that when the pixel unit vectors are used as directions along which to trace our rays, every point in the simulation can automatically be associated with a ray.
        In our simulations we use \textsc{HEALPix}-order 5, which corresponds to $n_\mathrm{rays} = 12 288$.

\section{Single-star models}\label{sec:single}
    
        \begin{table}
            \caption{Model parameters for the single star simulation.}
            \label{tab:paramSingle}
            \begin{center}
            \begin{tabular}{c c c}
                \hline\hline
                Parameter & Value & Unit \\
                \hline
                $\dot M_\mathrm{AGB}$ & $10^{-8}$ or $3 \times 10^{-6}$ & ${\rm M}_\odot \, {\rm yr}^{-1}$\\
                $M_\mathrm{AGB}$ & 1.02 & M$_\odot$\\
                $L_\mathrm{AGB}$ & 4384 & L$_\odot$\\
                $T_{\rm eff,AGB}$ & 2874 & K\\
                $R_\mathrm{AGB}$ & 1.24 & au\\
                \hline
                $R_{\rm inj}$ & 1.24 & au\\
                $v_{\rm inj}$ & $33$ or $25.2$ & ${\rm km\,s}^{-1}$\\
                \hline
                $\gamma$ & 1.2 & \\
                $\mu$ & 2.381 & \\
                \hline
            \end{tabular}
            \end{center}
            {\textbf{Notes.} \footnotesize{$\dot M_\mathrm{AGB}$ is the mass-loss rate of the AGB star, $M_\mathrm{AGB}$ its mass, $L_\mathrm{AGB}$ its luminosity, $T_{\rm eff,AGB}$ its surface temperature, and $R_\mathrm{AGB}$ its radius. $R_{\rm inj}$ is the wind injection radius and $v_{\rm inj}$ the initial injection velocity \citep[See][for details about the wind injection properties]{Siess2022}. $\gamma$ is the adiabatic index and $\mu$ the mean molecular weight of the gas.}}
        \end{table}
        
        \begin{figure}
            \centering
            \includegraphics[width=\linewidth]{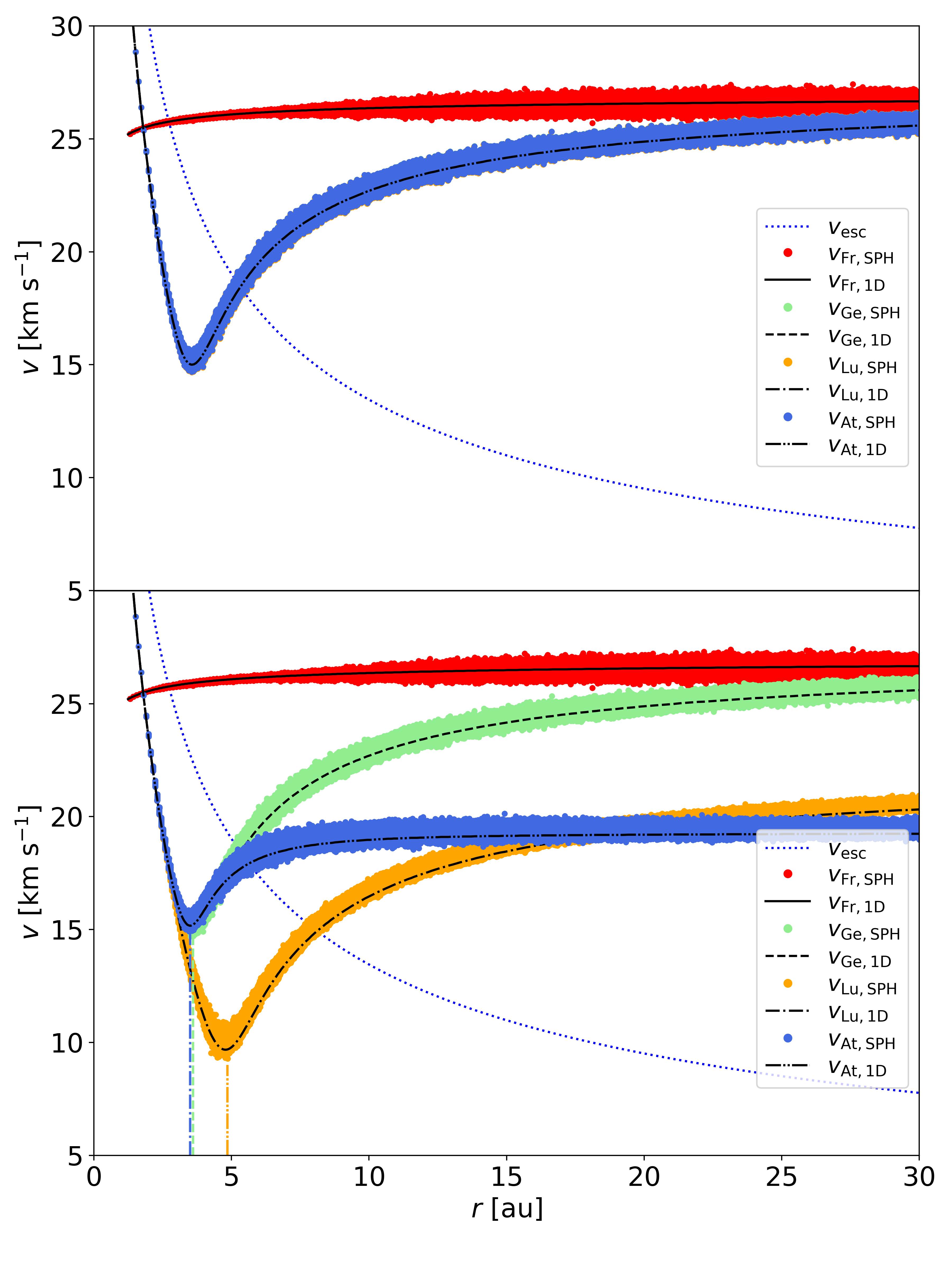}
            \caption{Velocity profiles from the SPH simulation (points) and 1D semi-analytical wind solution (lines) in a single-star configuration for the low ($\dot M_\mathrm{AGB} = 10^{-8} \, {\rm M}_\odot \, {\rm yr}^{-1}$, upper panel) and the high ($\dot M_\mathrm{AGB} = 3 \times 10^{-6} \, {\rm M}_\odot \, {\rm yr}^{-1}$, bottom panel)  mass-loss rate models. The free-wind approximation ({red} SPH points and 1D solid line profile), the geometrical approximation ({light-green} and dashed line), the Lucy approximation ({orange} and dash-dotted line), and the attenuation approximation ({blue} and dashed-double dotted line) are shown. The geometrical, Lucy, and attenuation approximation overlap in the upper panel. The vertical lines in the lower panel indicate the corresponding dust condensation radius for the geometrical, Lucy, and attenuation approximation. The dotted line shows the escape velocity.}
            \label{Fig:VelSingle}
        \end{figure}
        
        \begin{figure}
            \centering
            \includegraphics[width=\linewidth]{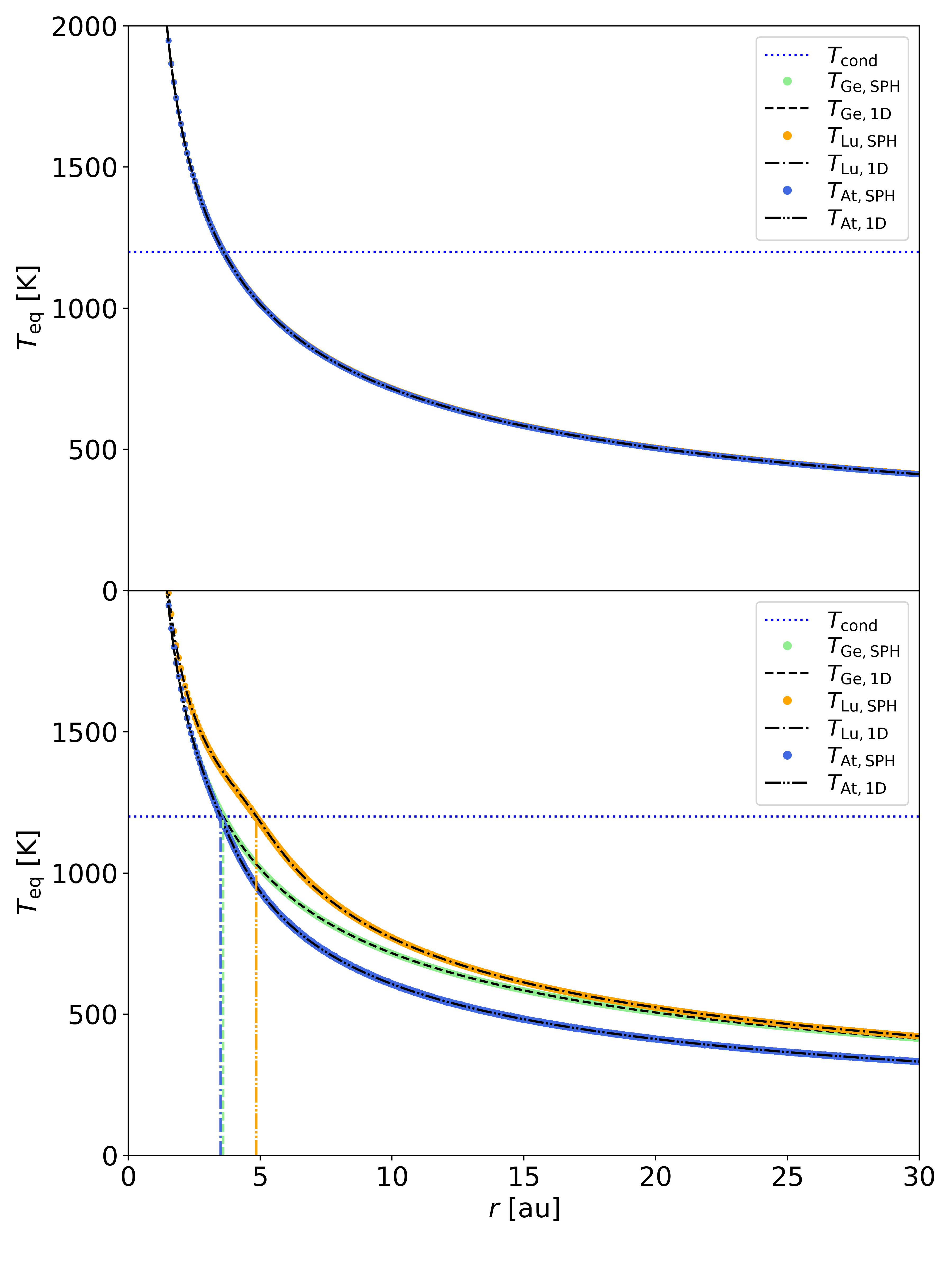}
            \caption{Radiative equilibrium temperature of the SPH particles (points) and 1D  semi-analytical wind solutions (lines) in a single-star configuration for the low mass-loss  ($\dot M_\mathrm{AGB} = 10^{-8} \, {\rm M}_\odot \, {\rm yr}^{-1}$, upper panel) and the high ($\dot M_\mathrm{AGB} = 3 \times 10^{-6} \, {\rm M}_\odot \, {\rm yr}^{-1}$, bottom panel) mass loss-rates models. The free-wind approximation ({red} SPH points and 1D solid line profile), the geometrical approximation ({light-green} and dashed line), the Lucy approximation ({orange} and dash-dotted line), and the attenuation approximation ({blue} and dashed-double dotted line) are shown. The geometrical, Lucy, and attenuation approximation overlap in the upper plot. The vertical lines in the lower panel indicate the corresponding dust condensation radius for the geometrical, Lucy, and attenuation approximation.}
            \label{Fig:TdustSingle}
        \end{figure}
        
    To investigate the impact of each of the four radiative transfer prescriptions on the outflow velocity profile, we perform eight single star simulations adopting a low ($10^{-8}\, {\rm M}_\odot \, {\rm yr}^{-1}$) and  high ($3 \times 10^{-6}\, {\rm M}_\odot \, {\rm yr}^{-1}$) mass-loss rate.
    These mass-loss rates roughly cover the range observed in AGB stars \citep{Habing2004, Ramstedt2008}. Higher mass-loss rates where not considered, because in the Lucy and attenuation approximations a wind could not be launched with these parameters.
    The properties of the models are listed in Table \ref{tab:paramSingle}, where we adopt the stellar parameters from \cite{Chen2020}.
    In the geometrical, Lucy, and attenuation approximation, the wind injection velocity is set to $v_{\rm inj} = 33 \, {\rm km \, s}^{-1}$. With this initial velocity, the particles can reach the dust condensation radius, which is a necessary condition to launch a wind, and have reasonable terminal velocities in agreement with observations.
    For the free-wind approximation, which  does not take into account dust condensation, an injection velocity of $v_{\rm inj} = 25.2 \, {\rm km \, s}^{-1}$ is chosen to match the terminal wind speed of the other cases.

    \subsection{Low mass-loss rate}
        The velocity profiles obtained in the low mass-loss rate regime ($\dot M_\mathrm{AGB} = 10^{-8} \, {\rm M}_\odot \, {\rm yr}^{-1}$) are displayed in the upper panel of Fig.~\ref{Fig:VelSingle}. In the free-wind approximation (red), the material is immediately accelerated outwards, and propagates at almost constant velocity. As $\Gamma$ is set equal to one, no external force is applied, and only the pressure gradient close to the star influences the wind velocity. For the geometrical approximation (light-green),  close to the wind injection zone, the material is too hot to condensate into dust (see Fig.~\ref{Fig:TdustSingle}, upper panel) and without the radiative acceleration, the gas pressure gradient is insufficient to drive the wind. Therefore, the velocity of the material initially decreases until it reaches the dust condensation radius (where $T_\mathrm{eq} = T_{\rm{cond}}$ at $R_{\rm Ge,dust}$ = 3.6~au, see Fig.~\ref{Fig:TdustSingle} upper panel). At this point, dust forms and the radiation force on the particles accelerates the material outwards. This results in a different velocity profile than the free-wind approximation. The velocity profiles in the Lucy and attenuation approximations (orange and blue) look identical to that of the geometrical approximation, because in the low mass-loss regime, densities in the wind are low. As a consequence, the optical depths $\tau_L$ (Eq.~\ref{eq:tauL})
        and $\tau$ (Eq.~\ref{eq:tau}) are very small, such that their effects are negligible.
    
    \subsection{High mass-loss rate}\label{sec:highmassloss}
    
        \begin{figure*}
            \centering
            \includegraphics[width=.95\linewidth]{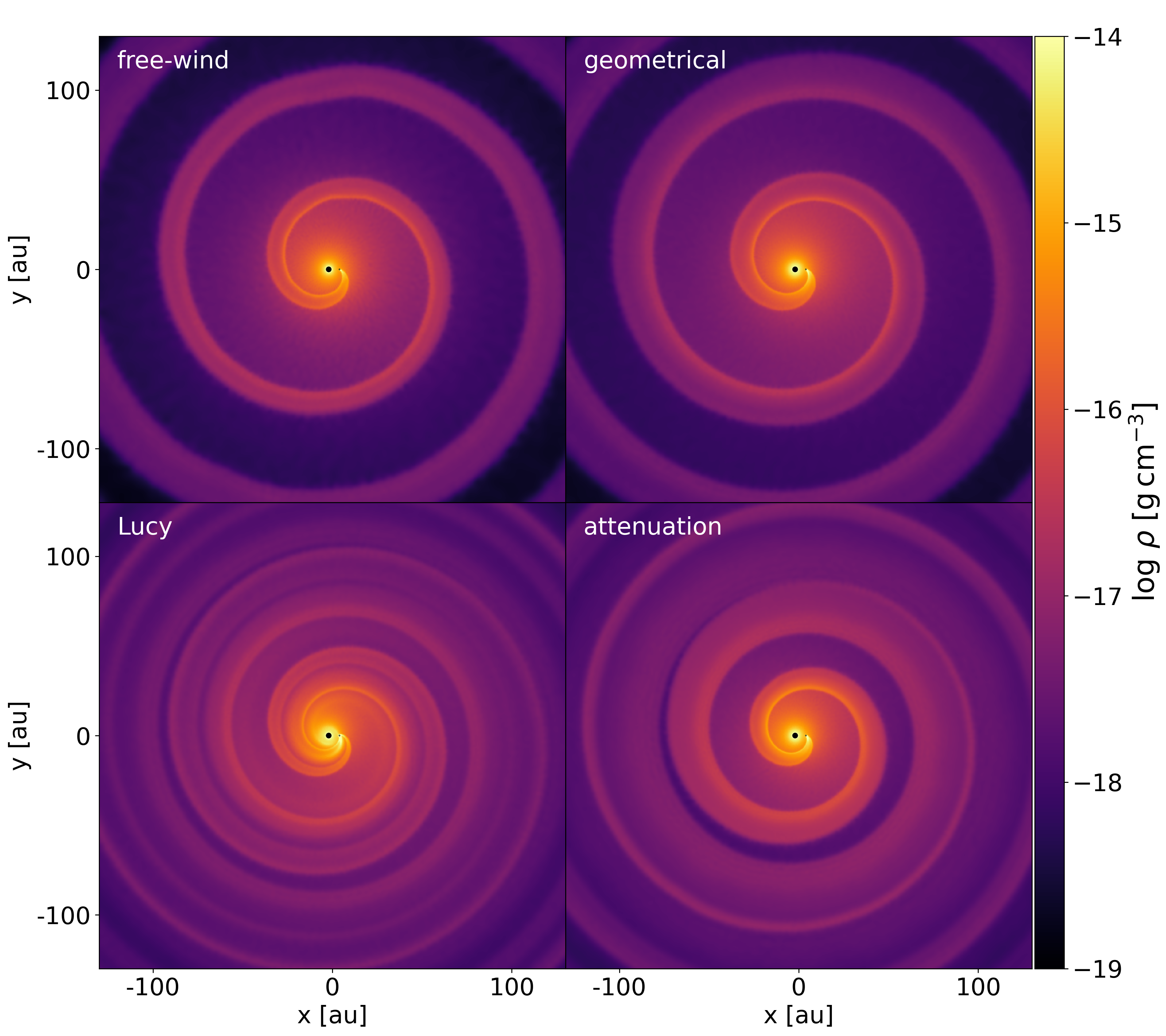}
            \caption{Density distributions in a slice through the orbital plane for the four simulations, each of which using different radiative transfer prescription: free-wind (top left), geometrical (top right), Lucy (bottom left), and attenuation (bottom right) approximation, for the high mass-loss rate case with a binary companion. Both stars are on the x-axis, where the primary AGB star is on the left, and the companion on the right.}
            \label{Fig:2Ddensity}
        \end{figure*}
        \begin{figure*}
            \centering
            \includegraphics[width=.95\linewidth]{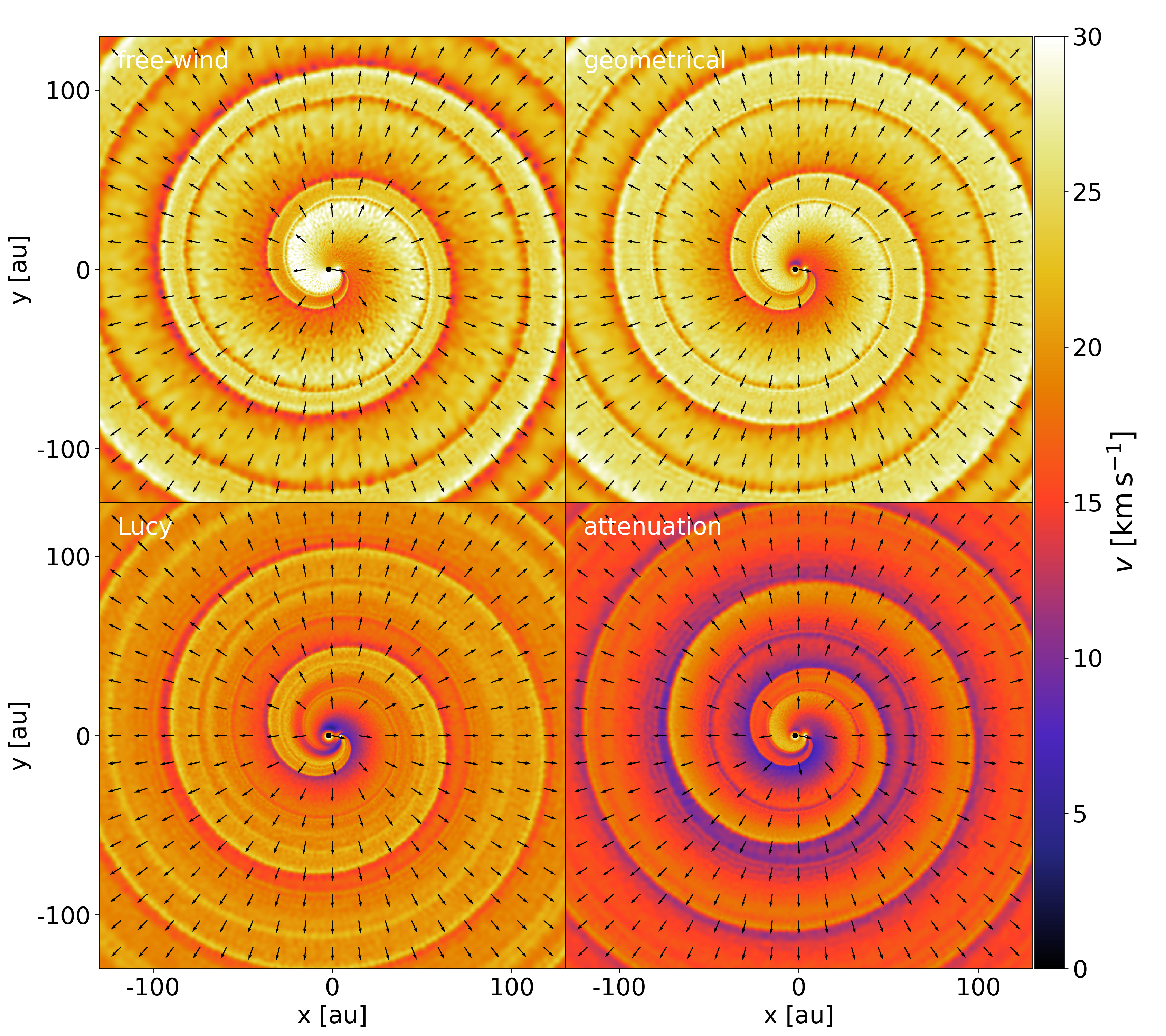}
            \caption{Same as Fig.~\ref{Fig:2Ddensity}, but for the velocity distribution.}
            \label{Fig:2Dspeed}
        \end{figure*}

        \begin{figure*}
            \centering
            \includegraphics[width=.76\linewidth]{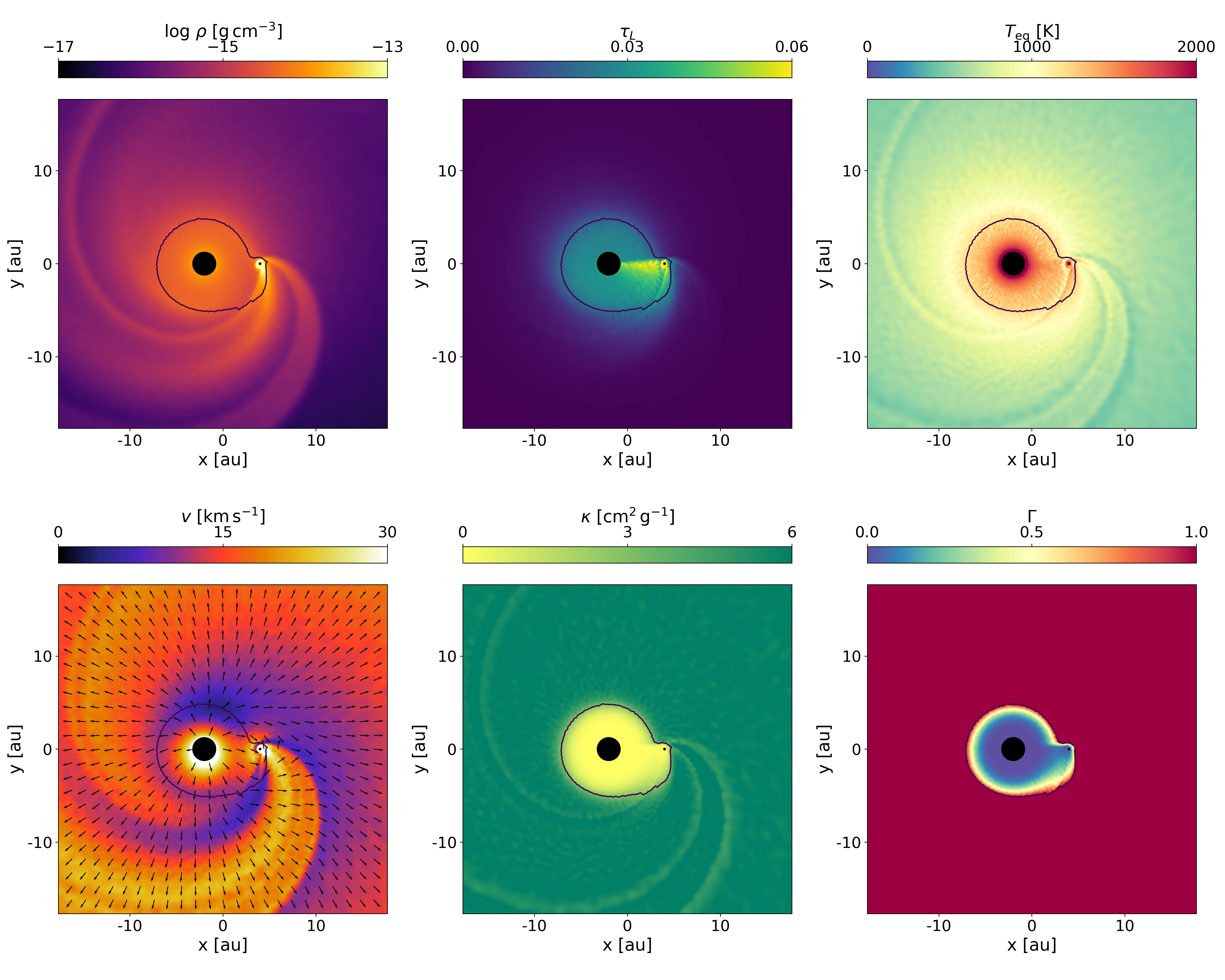}
            \caption{Relevant properties of the high mass-loss rate binary simulation using the Lucy approximation. The density is plotted in the upper left panel, the Lucy optical depth in the upper middle panel, the dust temperature in the top right panel, the velocity in the lower left panel, the opacity in the lower middle panel, and the Eddington factor in the lower right panel. The thin solid black contour indicates the location of the dust condensation surface.}
            \label{Fig:2DplotLucy}
            \centering
            \includegraphics[width=.76\linewidth]{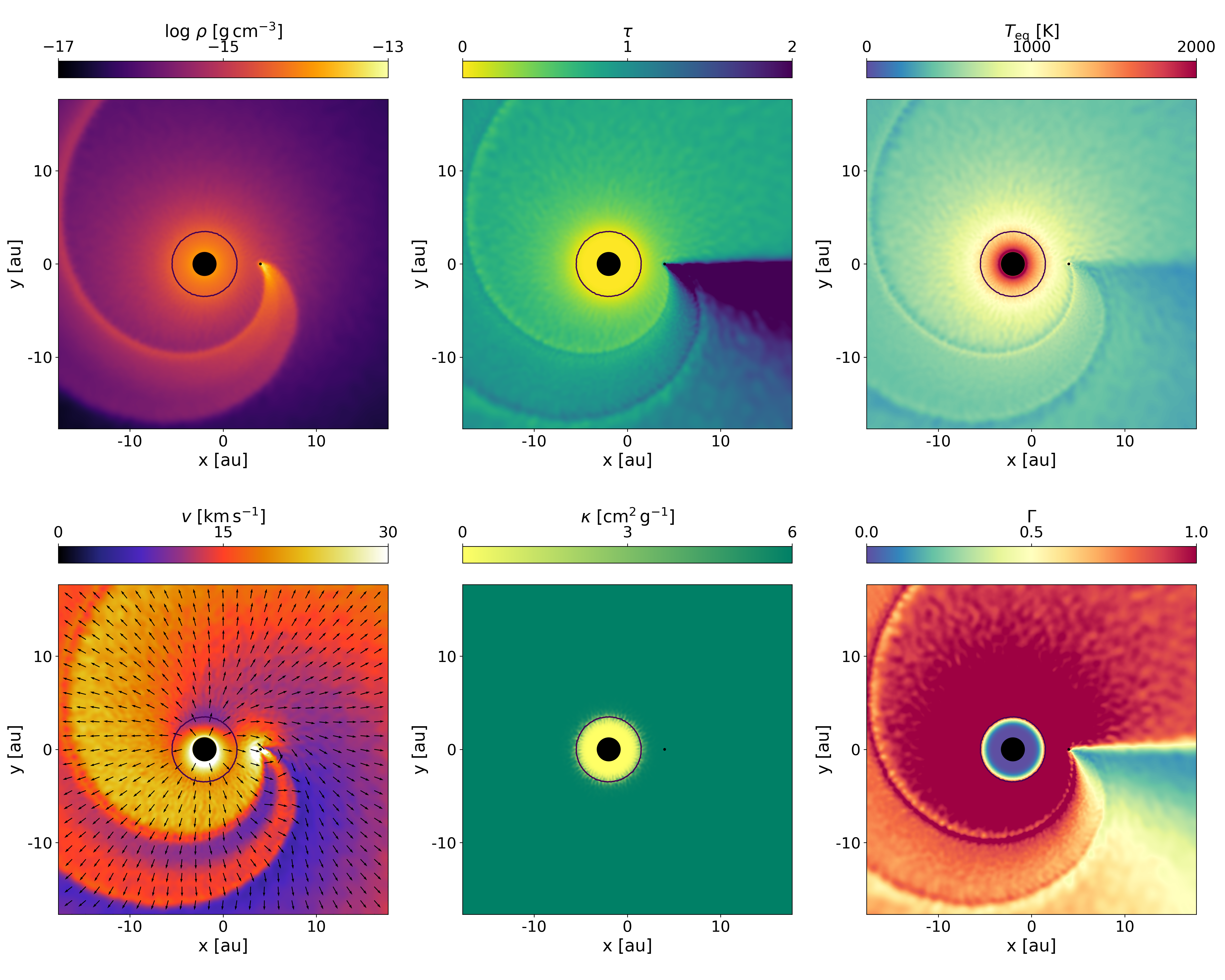}
            \caption{Same as Fig.~\ref{Fig:2DplotLucy} but for the attenuation approximation. $\tau_L$ is replaced by $\tau$ in the upper middle panel.}
            \label{Fig:2DplotAtten}
        \end{figure*}
        
        The velocity profiles obtained in the high mass-loss rate regime ($\dot M_\mathrm{AGB} = 3 \times 10^{-6} \, {\rm M}_\odot \, {\rm yr}^{-1}$) can be seen in the bottom panel of Fig.~\ref{Fig:VelSingle}. In the free-wind and geometrical approximation (red and light-green), the equations of motion (Eqs.~\ref{eq:consofmass}, \ref{eq:consofmom} and \ref{eq:consofener}) are independent of the mass-loss rate, even with the Bowen cooling prescription (Eq.~\ref{eq:bowen_cooling}) activated. However, this conclusion does not hold anymore if HI cooling is considered, because of the non-linear dependence of this rate on the density.
        In these single-star models, no shock waves are present, the temperature remains below $3000$~K, and cooling due to HI is inefficient.
        Therefore, the velocity profiles in the free-wind and geometrical approximation appear to be identical to the low mass-loss rate case (upper panel of Fig.~\ref{Fig:VelSingle}). However, this is not the case for the Lucy and the attenuation approximation (orange and blue), because the density is sufficiently high that $\tau_L$ (Eq.~\ref{eq:tauL})
        and $\tau$ (Eq.~\ref{eq:tau}) become non-negligible, or even larger than the dilution factor $W(r)$. This directly impacts the dust temperature profiles (see Fig.~\ref{Fig:TdustSingle}, bottom panel) and modifies the dust condensation radii (vertical lines in Figs.~\ref{Fig:VelSingle} and \ref{Fig:TdustSingle}). An illustration of the optical depths profiles is presented in Figs.~\ref{fig:1DProfile_tau_lucy} and \ref{fig:1DProfile_tau}.
        
        In the Lucy approximation, the dust temperature is increased compared to the geometrical case (Eq.~\ref{eq:TeqLucy}). Although $\tau_L$ is small, it is multiplied by $T_\star^4$, which can make its contribution significant. The condensation temperature is thus reached further out from the AGB star, at $R_{\rm Lu,dust} = 4.85$~au (Fig.~\ref{Fig:TdustSingle}).
        For the attenuation approximation, the temperature decreases rapidly, owing to the $e^{-\tau}$ factor (Eq. \ref{eq:TeqAtten}), and as a consequence, dust forms closer to the AGB star (at  $R_{\rm Ge,dust} = 3.50$~au). Thus, the radiation force is activated at higher velocities compared to the Lucy simulation. 
        Once the wind material has passed the dust condensation radius, the radiation force becomes independent of the approximation, used to determine $T_\mathrm{eq}$, since the dust opacity is now almost constant ($\kappa_d \approx \kappa_\mathrm{max}$). Beyond the condensation radius, the shape of the velocity profile in the Lucy approximation is similar to that of the geometrical approximation, but with a lower asymptotic value. 
        This is different in the attenuation approximation. Here, not only does the dust temperature decrease exponentially with $\tau$, but so does the radiation force (Eq.~\ref{eq:FAtten}). Thus, the acceleration of the material drops more rapidly, and the terminal wind velocity reaches lower velocities, yielding a flatter velocity profile in this case.

\section{Binary-star models}\label{sec:binary}

    An interesting application of the present implementation is studying the impact of a companion star or planet on the dynamics and morphology of the AGB wind. It is important that we investigate the applicability and differences of the four approximations in three-dimensional, non-spherically symmetric models, in which a binary companion is perturbing the outflow.
    We used the same setup as before, but add a companion with mass $M_\text{comp} = 0.51 \, {\rm M}_\odot$ and an accretion radius $R_\text{acc} = 0.1$~au at an orbital separation of 6~au in a circular orbit (adopted from \citealp{Chen2020}). The simulations are evolved for a total of six orbital periods, reaching a quasi-steady state.
    As explained in Sect. \ref{sec:single}, the Lucy and attenuation approximation give almost identical results as the geometrical approximation in the low mass-loss rate regime. Further, for the geometrical and free-wind approximation, the effect of changing the mass-loss rate is weak, hence we limit this discussion to the high mass-loss rate regime. 
    
    \subsection{Wind-companion interaction strength}\label{sec:epsilon}
        \cite{Maes2021} found that the ratio of the energy density of the companion to the kinetic energy of the wind gives a good indication of the complexity of the outflow and magnitude of the wind-companion interaction. This parameter is defined as 
        \begin{equation}
            \label{vareps}
            \varepsilon = \frac{e_{\text{grav}}}{e_{\text{kin}}} = \frac{(24 G^3 M_\text{comp}^2 M_{\text{AGB}})^{1/3} } {v_\text{w}^2 a (1-e)},
        \end{equation}
        where $v_\text{w}$ is an estimate of the wind velocity at the location of the companion, which is calculated as
        \begin{equation}
            v_\text{w} = \sqrt{v_\text{single}^2(r = a) + v_\text{AGB}^2},
        \end{equation}
        with $v_\text{single}(r)$ the wind velocity at radius $r$ in the corresponding single-star model, $v_\text{AGB}$ the orbital velocity of the AGB star, and  $e$ the eccentricity (here zero).
        The higher $\varepsilon$, the stronger the interaction of the companion with the wind, and the more complex the resulting outflow is expected to be.
        The only effect of our different approximations on this $\varepsilon$ value, is the change in $v_\text{single}(r=6 \, {\rm au})$, for which the four different values can be read from the bottom panel of Fig.~\ref{Fig:VelSingle}.
        This velocity is $\sim 26, 20, 18, \, {\rm and} \, 12 \, {\rm km \, s}^{-1}$ in case of the free-wind, geometrical, attenuation, and Lucy approximation, respectively. This results in $\varepsilon$ values of $0.4, 0.7, 0.8, \, {\rm and} \, 1.7$, respectively.
        This indicates that the interaction of the companion with the wind is expected to be the weakest for the free-wind approximation, and strongest for the Lucy approximation.

    \subsection{Wind structures}\label{sec:WindStructures}
        To illustrate the morphology and wind structures that result from the wind-companion interaction, Figs.~\ref{Fig:2Ddensity} and \ref{Fig:2Dspeed} display the density and velocity maps, respectively, in slices through the orbital plane for the four approximations. The density profile in the meridional plane can be found in Fig.~\ref{Fig:2DdensityMeridional}. These figures reveal that the wind structures and global morphologies depend sensitively on the treatment of the radiative transfer.
        
        In the free-wind approximation, the relatively weak wind-companion interaction strength (low $\varepsilon < 1$), creates a two-edged spiral structure attached to the companion, that shapes the wind into an approximate Archimedean spiral, as can be seen in the density profile in the orbital plane (Fig.~\ref{Fig:2Ddensity}, upper left). The creation of such Archimedean spirals is well studied and explained in detail by, for example,  \cite{Malfait2021} and \cite{Maes2021}. Due to the high wind velocity with respect to the orbital velocity, this thin high-density spiral propagates rapidly outwards with a radial velocity of about $27 \, {\rm km \, s}^{-1}$, which results in relatively wide inter-arm low-density gaps, with a width approximately equal to the distance travelled by the spiral structure in one orbital period (see Fig.~\ref{Fig:2Dspeed}, upper left). The meridional plane density distribution (Fig.~\ref{Fig:2DdensityMeridional}, upper left) shows that in the edge-on view, these spirals appear as thin, concentric arcs.
        
        In the geometrical and attenuation approximation, the wind structure close to the stars is similar to the free-wind approximation, with a two-edged spiral structure attached to the companion (upper and lower right panels in Fig.~\ref{Fig:2Ddensity}). Although the wind velocity around the companion, and thereby the $\varepsilon$ value, is similar for these approximations, the morphology of these systems is different. As the velocity in the geometrical approximation accelerates up to about $\sim 27 \, {\rm km \, s}^{-1}$, reaching the same terminal velocity as the free-wind model (see Figs.~\ref{Fig:VelSingle} and \ref{Fig:2Dspeed}), the 2-edged spiral remains again relatively thin, and the inter-arm separation large. 
        
        This is not the case for the attenuation approximation, where the wind material in the high-density spirals only reaches a velocity of $\sim 20 \, {\rm km \, s}^{-1}$, and the low-density material in between the spirals has a velocity of 
        $\sim 10-15 \, {\rm km \, s}^{-1}$ (Fig.~\ref{Fig:2Dspeed}). This makes that the spiral structure more compressed, with smaller low-density inter-arm gaps. Because there is a velocity dispersion within the spiral, the outer frontward spiral edge has a higher radial velocity than the inner backward spiral edge, such that after $\sim 1.75$ orbital periods the outer spiral edge catches up and interacts with the previous inner spiral edge, that originated one orbital period earlier (around $x = 0$~au, $y = 80$~au). After this interaction, one approximate Archimedean spiral structure remains, and the inter-arm low-density gaps disappear \citep{Malfait2021, Maes2021}.
        In the meridional plane, the spiral appears again as arcs (Fig.~\ref{Fig:2DdensityMeridional}, lower right). This plot also shows that the high-density structure is more compressed, with smaller low-density inter-arc regions, and that the outer edge of the widening arcs catches up and interacts with the previous inner edge (overlap first visible at $x = 100$~au, $y = 0$~au).

        Due to the larger dust condensation radius in the Lucy approximation, the wind velocity around the location of the companion is significantly lower than in the previously discussed models (bottom panel in Fig.~\ref{Fig:VelSingle}, at $r=6$~au, and $r<10$~au region in Fig.~\ref{Fig:2Dspeed}). This allows the companion to compress more wind material around it, such that instead of a 2-edged spiral structure, there is one spiral originating behind the companion, and a second 2-edged bow shock spiral originating in front of the companion \citep{Malfait2021, Maes2021}. This can be seen in more detail in the zoomed-in density plot in the upper left plane of Fig.~\ref{Fig:2DplotLucy}.
        Moreover, due to the strong compression of gas around the companion, as well as sufficient cooling to reduce the thermal pressure, an accretion disk has formed. 
        In the Lucy approximation, there is no radiative force active close to the companion (see Sect.~\ref{sec:radforce}), so the forming accretion disk is not blown away, facilitating its formation.
        This accretion disk is shown in more detail in Fig.~\ref{fig:AccrDisk}, where the density distribution is overplotted with velocity vectors. Fig.~\ref{fig:AccrDisk} displays how material spirals in towards the companion sink particle through a high-density disk. 
        For a more elaborate description of accretion disks, see \cite{Lee2022} and Malfait et al.\ (in prep). 
        In the meridional plane, the bow shock spiral translates into an expansion of the edge-on arcs (Fig.~\ref{Fig:2DdensityMeridional}, lower right).
        
        \cite{Chen2020} used the attenuation approximation in their simulations, and report the formation of both an accretion and a circumbinary disk. The absence of these features in our computations stems from the fact that the terminal wind velocity and cooling prescriptions are different between these two works. In our simulations, the terminal velocity is $\sim 20 \, {\rm km \, s}^{-1}$, higher than the value of $\sim 15 \, {\rm km \, s}^{-1}$ found by \cite{Chen2020}. With a faster wind, the particles interact less with the companion (Sect.~\ref{sec:epsilon}) and pass over the companion preventing the formation of the disk.
        The disk may, however, become visible when decreasing the accretion radius of the companion \citep{Lee2022}.
        The same is true for the circumbinary disk, as a higher terminal wind velocity prevents the formation of such structures. 
        \cite{Chen2020} also include molecular cooling, associated with H$_2$, H$_2$O, and CO, which contribute to reduce the heat (pressure) and to provide more favourable conditions for gas condensation. These processes mostly influence the formation of circumbinary disks, since in a circumstellar accretion disk, the temperature is higher, and atomic cooling is expected to dominate (\citealp{Mastrodemos1999}; see also \citealp{Woitke1996} for information on the dominant cooling processes).
    
    \subsection{Impact of the approximations on the radiation force}\label{sec:radforce}
        
        In the free-wind approximation, the radiation force is not explicitly calculated, but is assumed to be equal to the gravitational force of the AGB star. For the geometrical approximation, its expression does not depend on the presence of a companion (Eqs.~\ref{eq:GammaGeometrical} and \ref{eq:TdustGeometrical}). This is, however, not the case for the Lucy and attenuation approximation, because of the directional dependence of $\tau_L$ (Eq~.\ref{eq:tauL}) and $\tau$ (Eq.~\ref{eq:tau}), that enter the evaluation of $T_\mathrm{eq}$ and $J$. The density distribution $\rho$, optical depths ($\tau$ and $\tau_L$), dust temperature $T_{\rm{eq}}$, wind velocity $v$, opacity $\kappa=\kappa_d+\kappa_g$, and Eddington factor $\Gamma$ in the orbital plane are shown in Figs.~\ref{Fig:2DplotLucy} and \ref{Fig:2DplotAtten} for the Lucy and attenuation approximations, respectively.
        
        In the Lucy simulation, close to the companion, inside the 2-edged bow shock, the density and $\tau_L$ are high (upper middle panel). The mean intensity is thus increased in that region (Eq.~\ref{eq:J_lucy}) and under the condition of LTE, the equilibrium temperature is locally higher (Eq.~\ref{eq:Jdef}). This brakes the symmetry and the dust condensation surface, inside which no dust forms and which is defined as the region where $T(r) =T_\mathrm{cond}$, is not spherical anymore. This 3D surface is shown as the black 2D contour in the various panels. The opacity increases rapidly across this surface, which reflects directly on the Eddington factor $\Gamma$ (bottom right panel).
        
        These features are different in the simulation with the attenuation approximation (Fig.~\ref{Fig:2DplotAtten}). Because the dust condensation radius is significantly smaller than the orbital separation, the dust condensation surface remains approximately spherically symmetric, as shown by in the various panels.
        The effect of this attenuation is clearly visible behind the companion, where the optical depth starts to deviate from spherical symmetry. This effect can be seen as a shadow behind the companion ($x>4$~au side along $y = 0$~au). This shadow is cast as the material close to the companion forms a higher-density spiral structure,
        that absorbs the radiation from the AGB star. This effect also reduces the dust temperature in the region behind the companion (upper right plot). As the dust temperature is already sufficiently low for dust to form, this does not influence the opacity. But since the equation for the radiation force (described by $\Gamma$, see Eq.~\ref{eq:FAtten}) contains a factor $e^{-\tau}$, the radiation force completely vanishes behind the companion.

\section{Discussion}\label{sec:discussion}
    \subsection{Accuracy of the ray-tracing approximations}
        To gauge the accuracy and quantify the benefit of our ray-tracer implementation, we compare the dust temperatures $T_{\rm eq}$ and the radiation forces, obtained with our prescriptions, to the results obtained with the 3D ray-tracing radiative transfer solver \textsc{Magritte}.
        We focus only on the Lucy and attenuation approximation, as those are the only two prescriptions that leverage the newly implemented ray-tracer.
        To make this comparison, we take the final snapshots of their respective simulations (discussed in Sect. \ref{sec:radforce}), and use these as an input for \textsc{Magritte}.
        
        Although \textsc{Magritte} is usually advertised as a line radiative transfer code \citep{DeCeuster2020b, DeCeuster2020a}, we only use its core ray-tracer and solver functions here.
        More specifically, we supply it with the grey opacity (Eq.~\ref{eq:kappadust}) and emissivity $\eta = \kappa \rho J$ (with $J$ defined as in Eq. \ref{eq:Jdef}), and let it solve the radiative transfer equation (Eq.~\ref{eq:RT}) along 2\,700 uniformly distributed rays, originating from each SPH particle.
        From the resulting intensities along those rays, we then derive the mean intensity $J$ and flux $F$.
        However, the emissivity depends on the dust temperature, which itself depends on the radiation field.
        Therefore, to obtain a self-consistent solution between the dust temperatures and the radiation field, we need to compute them in an iterative way.
        Starting from the analytic radiative equilibrium temperature of the geometrical approximation (Eq.~\ref{eq:TdustGeometrical}), the dust opacities are computed (Eq.~\ref{eq:kappadust}), and then \textsc{Magritte} is used to compute the mean intensity $J$.
        From this, the dust temperatures can then be recomputed using (Eq.~\ref{eq:Jdef}). This process is repeated until the change in the dust temperatures becomes negligible (the resulting mean relative temperature differences in the final iteration is 0.02\%, and the maximal 2\%).
        To reduce the computational cost, only the SPH particles within a radius of 30~au are included.
        This should not alter the results, as densities (and thus interactions) are diluted significantly beyond this radius.
        
        \subsubsection{Lucy approximation}\label{sec:discussionLucy}
            \begin{figure}
                \centering
                \includegraphics[width = \linewidth]{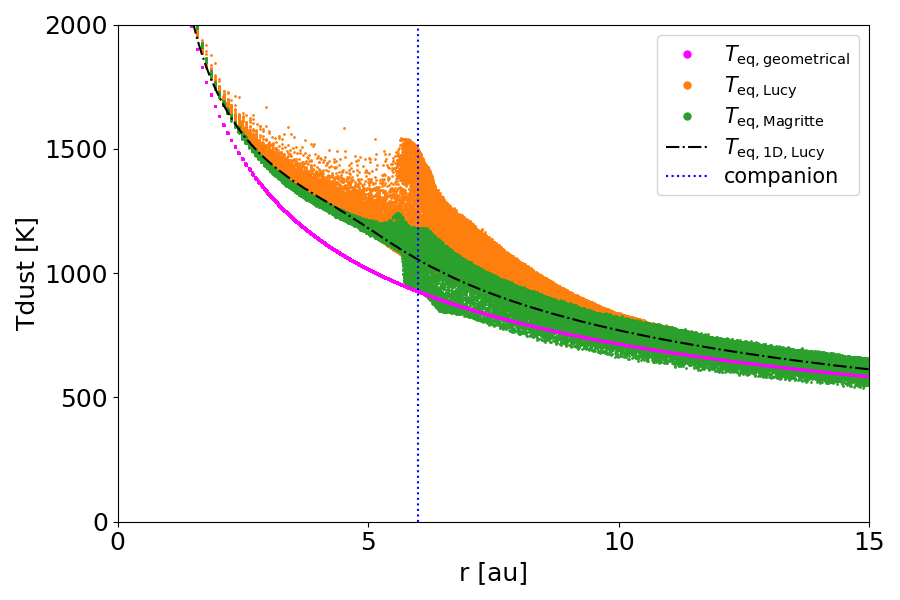}
                \caption{Radiative equilibrium temperature as a function of distance from the primary AGB star for the simulation using the Lucy approximation. Magenta represents the temperature calculated using the geometrical approximation, orange using the Lucy approximation, and green using \textsc{Magritte}. The lower envelope of the Lucy simulation (orange points) which is not visible as it falls behind the green points, closely follows the $T_\mathrm{eq,Lucy}$ dot-dashed line.}
                \label{fig:TdustLucyMatritte1D}
            \end{figure}
            \begin{figure}
                \centering
                \includegraphics[width = \linewidth]{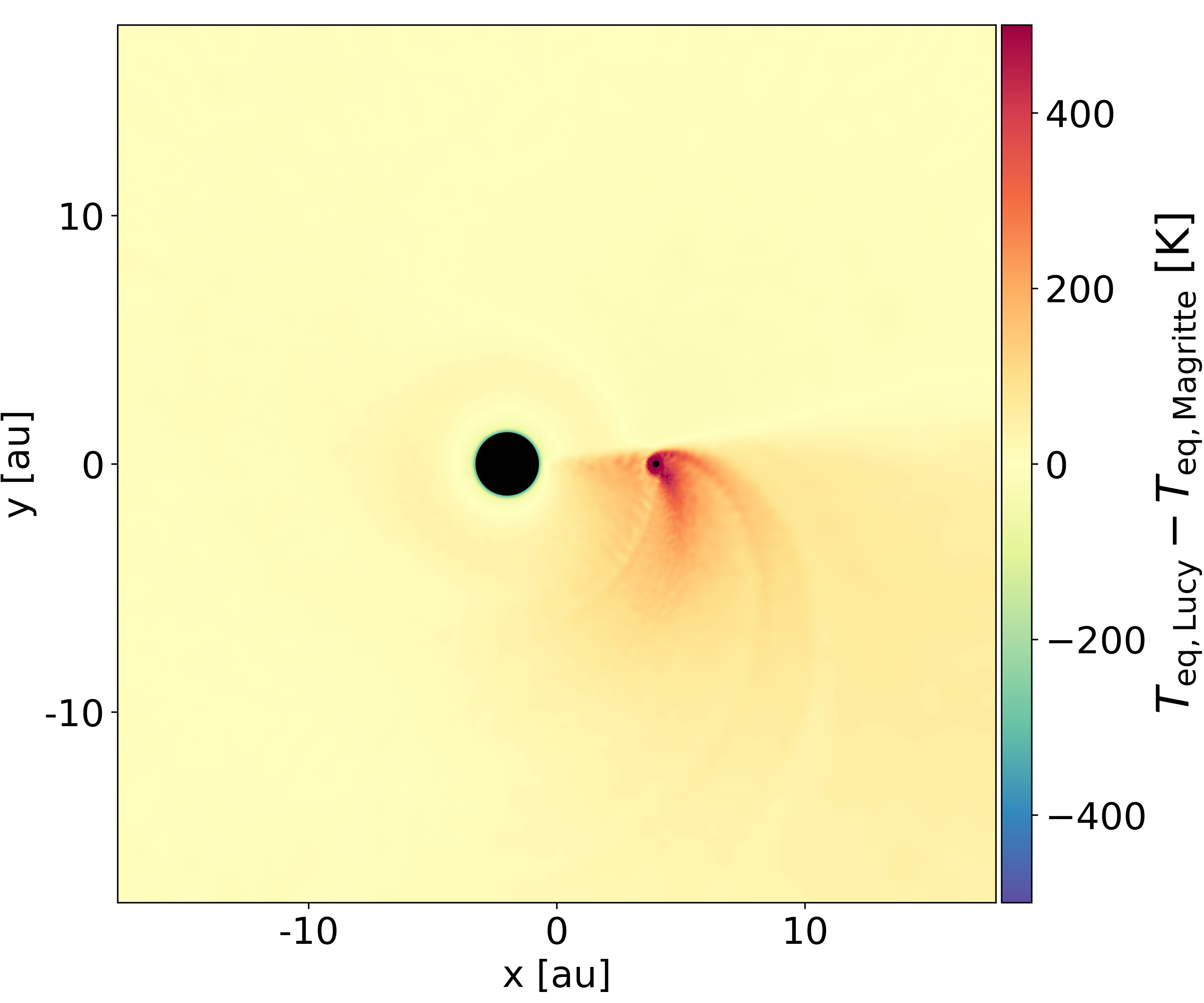}
                \caption{Difference in the orbital plane between the radiative equilibrium temperature calculated with the Lucy approximation and \textsc{Magritte}. The temperature is similar in the two cases except in the directions of the companion where the Lucy approximation yields higher temperatures.}
                \label{fig:TdustLucyMagritte2D}
            \end{figure}
            
            Fig.\ \ref{fig:TdustLucyMatritte1D} shows the dust temperature $T_{\rm eq}$, calculated using \textsc{Magritte}, as a function of distance from the AGB star.
            Here, the dust temperature in the geometrical approximation (used as initial temperature in \textsc{Magritte}) is shown in magenta, results obtained with the Lucy approximation (as calculated in \textsc{Phantom}) are shown in orange, and results from \textsc{Magritte} are shown in green. We see that inside the orbit ($r \la 6$~au), the 1D Lucy dust temperature nicely follows the \textsc{Magritte} prediction. Just before the location of the companion, a sudden drop in the \textsc{Magritte} dust temperature appears, as well as a slight increase at the top of the green curve (at $r = 5.5$ and 6~au).
            The increase coincides with the edge of the accretion disk and spiral arm, which are heated more efficiently.
            In the Lucy simulation, this effect is much more pronounced, due to the underlying assumption of spherical symmetry, such that, whenever a direction with high optical depth is encountered, an entire sphere at this optical depth is assumed.
            The particles from the \textsc{Magritte} post-processed model with a low dust temperature, are the particles located just behind the heated regions.
            The second drop in the dust temperature around $r\approx 6-6.3$~au occurs behind the companion, and is due to the bow shock acting as a shadow (see also Fig. \ref{fig:TdustLucyMagritte2D}).

            Fig.~\ref{fig:TdustLucyMagritte2D} shows the difference between the dust temperature computed in the Lucy approximation (Fig.~\ref{Fig:2DplotLucy}, upper right panel) and the one obtained with \textsc{Magritte}, for a slice through the orbital plane.
            While there are clear deviations from the Lucy temperature in the direction of the companion, in other directions, where the density profile is less perturbed, differences with respect to the Lucy temperature remain small.
            Hence, the Lucy approximation performs best in regions where the underlying assumption of spherical symmetry remains approximately valid. 
            In the direction of the companion, and especially in the accretion disk around the companion ($x = 4$~au, $y = 0$~au), the dust temperature is artificially heated, due to the underlying assumption of spherical symmetry, as explained above. Behind the companion, the shadow is not captured in the Lucy approximation, resulting in the deviations at $x>5$~au and $y<0$~au (red region).

            \begin{figure}
                \centering
                \includegraphics[width = \linewidth]{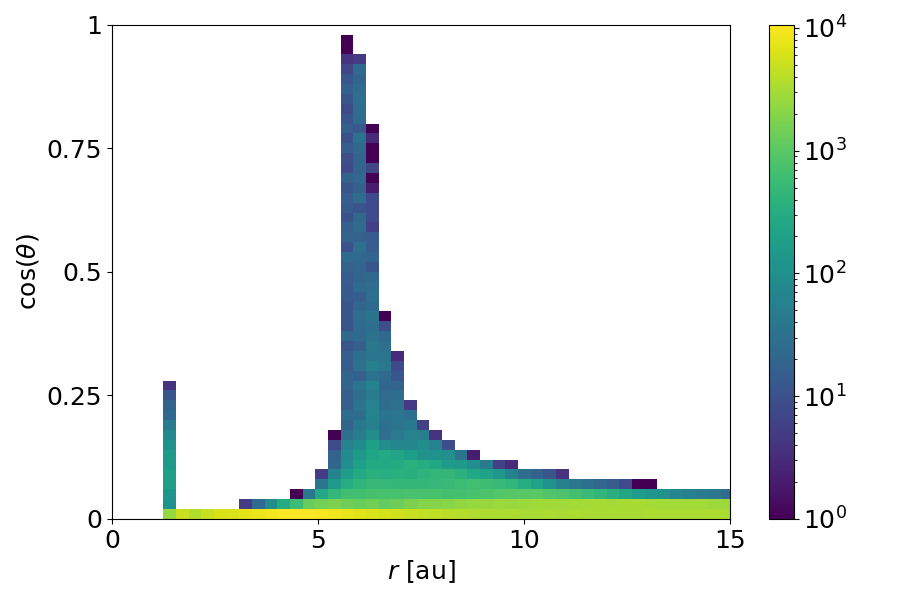}
                \caption{2D histogram of the relative non-radial component of the radiation force, as a function of distance from the AGB star, given by \textsc{Magritte} for the snapshot of the model using the Lucy prescription. The colourbar is in log-scale.}
                \label{fig:radcompLucy}
            \end{figure}
            \begin{figure}
                \centering
                \includegraphics[width = \linewidth]{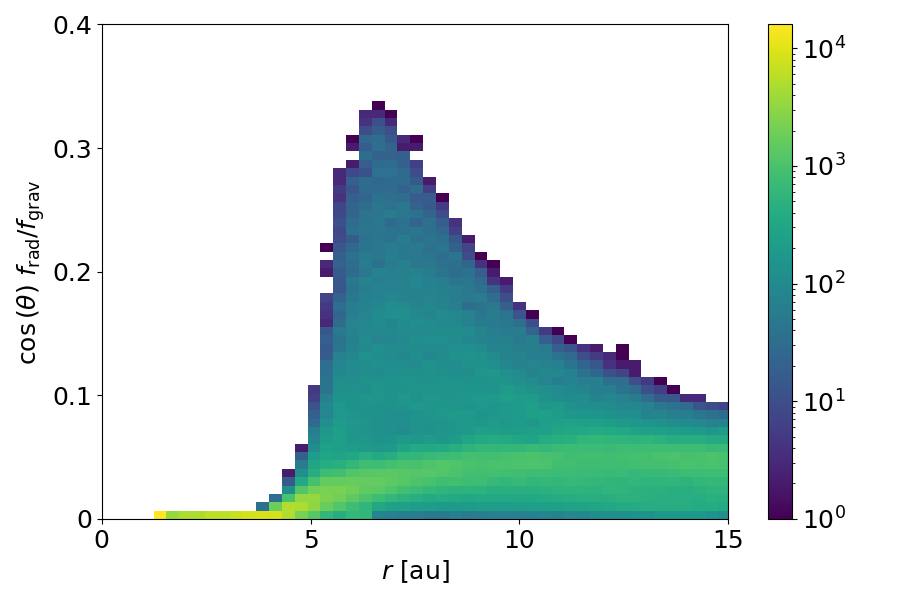}
                \caption{2D histogram of the non-radial component of the radiation force, relative to the gravitational attraction of the AGB star (see Eq.~\ref{eq:Delta}) as a function of distance from the AGB star, given by \textsc{Magritte} for the snapshot of the model using the Lucy prescription. The colourbar is in log-scale.}
                \label{fig:radcompscaledLucy}
            \end{figure}
            \begin{figure}
                \centering
                \includegraphics[width = \linewidth]{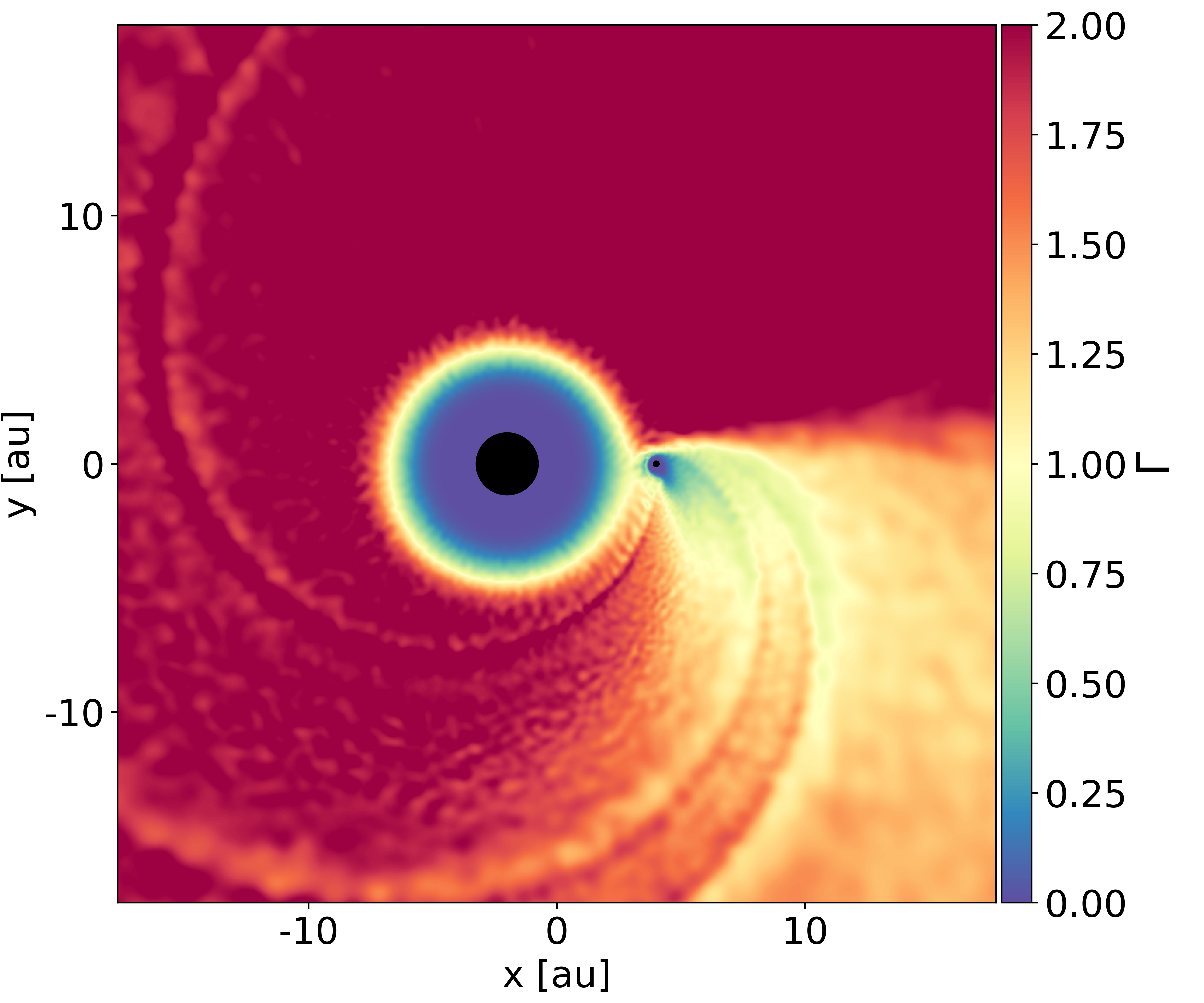}
                \caption{Eddington factor $\Gamma$ (Eq.~\ref{eq:Gamma}), calculated using the magnitude of the flux obtained with \textsc{Magritte},  in a slice through the orbital plane for the snapshot of the model, using the Lucy prescription.}
                \label{fig:radforceLucy2D}
            \end{figure}
            
            The analysis of the radiation force ($f_{\rm rad}$) is more complex, as this is a vector quantity.
            In all four prescriptions, the radiation force is assumed to be radial, thus pointing from the AGB star to the position of the SPH particle under consideration.
            To verify whether this is also true for the radiation flux computed with \textsc{Magritte}, the non-radial fraction of the radiation force ($\cos{(\theta)}$) can be computed
            \begin{equation}
            \cos{(\theta)} = \frac{\vec{f_{\rm rad}} \cdot \vec{r}}{||f_{\rm rad}||~||r||}\,.
            \end{equation}
            This quantity is shown as a function of distance from the AGB star in Fig.~\ref{fig:radcompLucy}.
            A clear spike is visible at the location of the companion ($r=6$~au), due to the dense accretion disk emitting a significant amount of radiation, which dominates the radiation force near the accretion disk.
            Although these forces are highly non-radial, they are relatively small in magnitude when compared to the local gravitational attraction of the AGB star.
            This quantity,
            \begin{equation}\label{eq:Delta}
                \frac{\cos{(\theta)} f_{\rm rad}}{f_{\rm grav}} =  
                \frac{\kappa(T_{\rm eq})\cos{(\theta)}||F||/c}{GM_{\rm AGB}/r^2}\,,
            \end{equation}
            is displayed in Fig.\ref{fig:radcompscaledLucy}. This shows that the assumption of a radial radiation force is reasonable.
            In the remainder of this analysis, we assume a radial radiation force and only consider its magnitude. The small spike around the inner boundary at $r=R_\mathrm{AGB}$ = 1.24~au is a numerical artefact of \textsc{Magritte} that reveals the discretization of the (spherical) stellar surface. This feature is also present in Fig.~\ref{fig:radcompscaledAtten}, but it is inconsequential.
            
            The Eddington factor $\Gamma$ (Eq.~\ref{eq:Gamma}) is shown for a slice through the orbital plane in Fig.~\ref{fig:radforceLucy2D}. 
            First, we compare $\Gamma$ resulting from \textsc{Magritte} (Fig.~\ref{fig:radforceLucy2D}) to $\Gamma$ resulting from the Lucy approximation (Fig.~\ref{Fig:2DplotLucy}, lower right panel).
            In the \textsc{Magritte} case, the dust condensation surface is a perfect sphere, and since it 
            is smaller than the orbital separation, it is not perturbed by the companion, and remains spherically symmetric.
            This is in contrast to the dust condensation surface in Fig.~\ref{Fig:2DplotLucy}, which is extended beyond the companion and even engulfs it.
            The radiation force inside the accretion disk around the companion is negligible ($\Gamma \sim 0$), since the high local densities make the radiation field isotropic, such that the contributions to the radiation force from different directions cancel each other out.
            Behind the companion ($x>5$~au and $y<0$~au), there is a low-$\Gamma$ region, and the resulting shadow looks similar to the shadow cast in that attenuation simulation (Fig.~\ref{Fig:2DplotAtten}, lower right panel).
            The radiation force increases again when moving further out, due to the fact that material above and below the shadow in the orbital plane shine and accelerate the material again.

        \subsubsection{Attenuation approximation}
            
            \begin{figure}
                \centering
                \includegraphics[width = \linewidth]{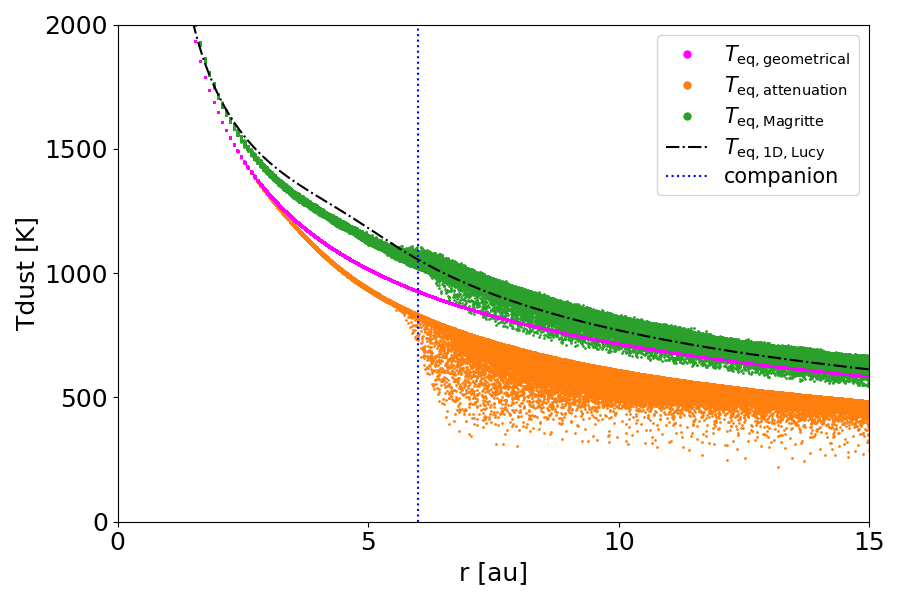}
                \caption{Radiative equilibrium temperature as a function of distance from the primary AGB star, for the simulation using the attenuation prescription. Magenta represents the temperature calculated, using the geometrical prescription, orange using the attenuation prescription, and green using \textsc{Magritte}.}
                \label{fig:TdustAttenMagritte1D}
            \end{figure}
            \begin{figure}
                \centering
                \includegraphics[width = \linewidth]{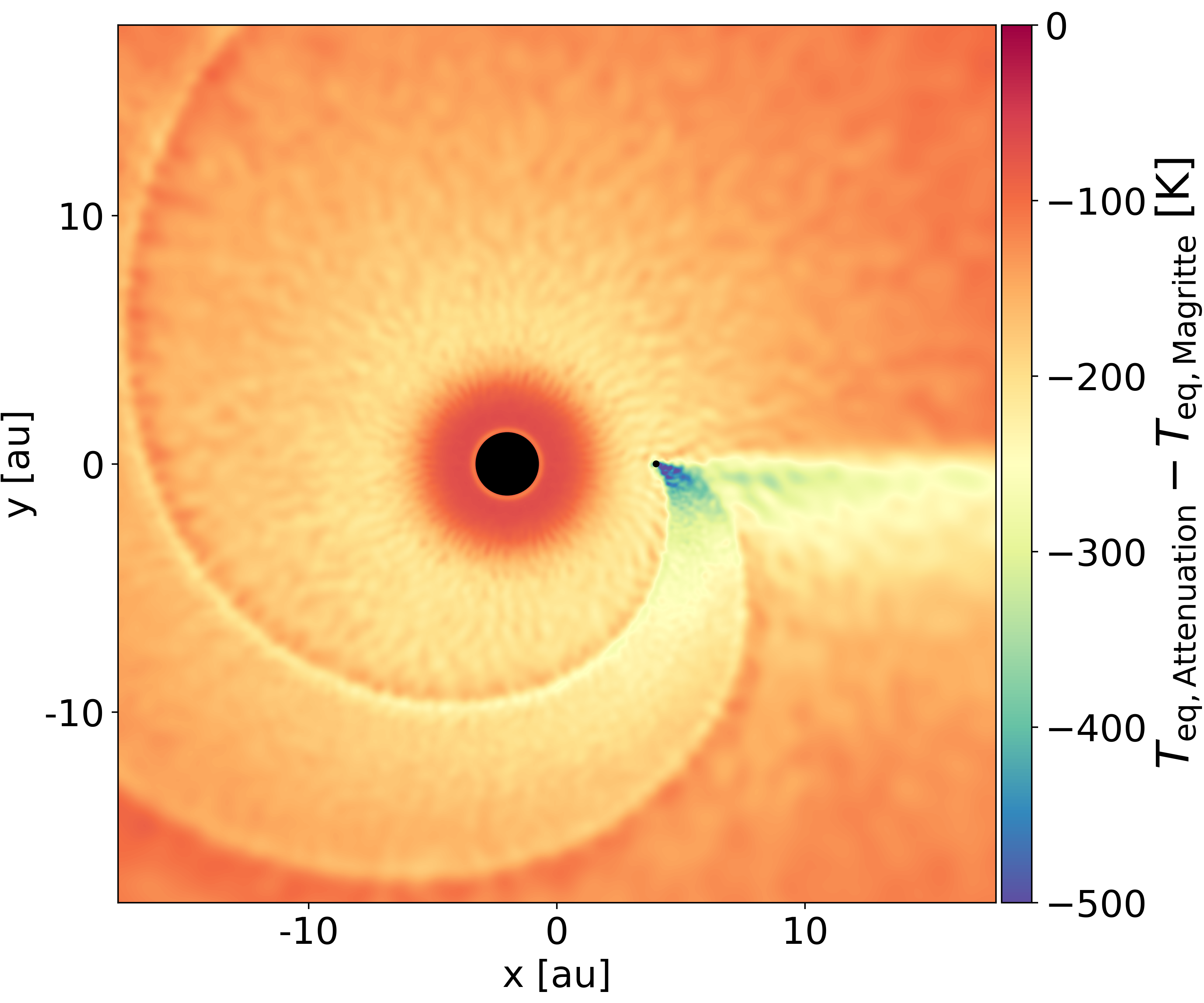}
                \caption{Difference in the orbital plane between the radiative equilibrium temperature, calculated with the attenuation prescription \textsc{Magritte}. Significant differences persist over the entire domain.}
                \label{fig:TdustAttenMagritte2D}
            \end{figure}
            
            Fig.~\ref{fig:TdustAttenMagritte1D} shows the dust temperature $T_{\rm eq}$, calculated using \textsc{Magritte} as a function of distance from the AGB star.
            The \textsc{Magritte} results for the snapshot of the simulation, using the attenuation prescription, show the same pattern as in the Lucy approximation. Within the orbit, the \textsc{Magritte} dust temperature follows closely the 1D Lucy approximation (and not the attenuation approximation), and at the location of the companion a drop appears due to the interaction of the high density spiral arm.
            In contrast to the Lucy approximation, there is only one drop and one peak, as there is only one spiral arm in this simulation, and no accretion disk or bow shock are formed. 
            In the attenuation approximation, the dust temperature decreases faster than in the geometrical and Lucy approximation, similar to the 1D profile (Fig.~\ref{Fig:TdustSingle}).
            This is caused by the fact that the attenuation approximation only accounts for absorption and not for re-emission, resulting in an underestimation of the intensities and dust temperatures.
            Looking at the region behind the companion, both \textsc{Magritte} and the attenuation approximation show a drop in the dust temperature.
            This drop starts at slightly lower radii in the attenuation approximation, compared to the \textsc{Magritte} calculation. In the attenuation approximation, the shadow immediately forms when high densities are encountered, as the optical depth increases. In a full radiative transfer treatment, when a high density region is encountered, the first layers absorb a lot of photons, locally trapping some of the radiation and producing a local heating that is seen in the peak in $T_\mathrm{eq, Magritte}$ at 6~au. After this peak, the temperature drops at $\approx 7$~au, as radiation escapes isotropically.
            The temperature decrease in the attenuation approximation is also too strong, and this is caused by the radiation being only blocked, while in \textsc{Magritte}, re-emission is accounted for.
            
            \begin{figure}
                \centering
                \includegraphics[width = \linewidth]{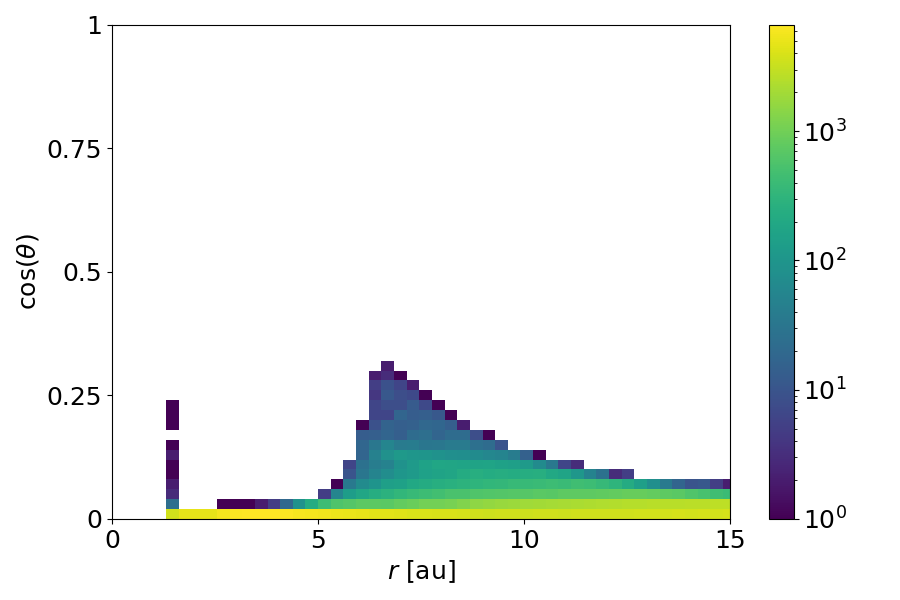}
                \caption{2D histogram of the relative non-radial component of the radiation force, as a function of the distance from the AGB star, given by \textsc{Magritte}, for the snapshot of the simulation using the attenuation prescription. The colourbar is in log-scale.}
                \label{fig:radcompAtten}
            \end{figure}
            \begin{figure}
                \centering
                \includegraphics[width = \linewidth]{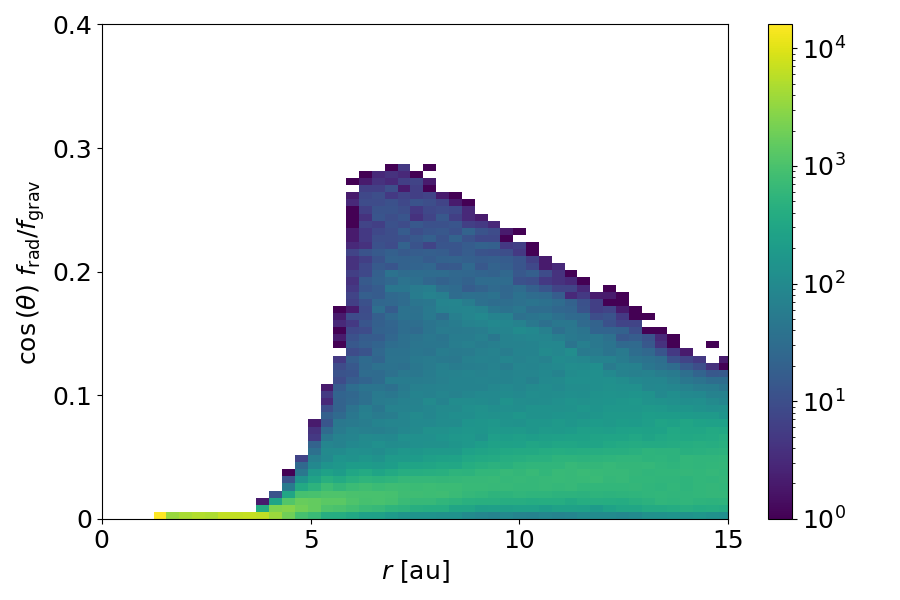}
                \caption{2D histogram of the non-radial component of the radiation force, relative to the gravitational attraction of the AGB star (see Eq.~\ref{eq:Delta}), as a function of distance from the AGB star, given by by \textsc{Magritte}, for the snapshot of the simulation using the attenuation prescription. The colourbar is in log-scale.}
                \label{fig:radcompscaledAtten}
            \end{figure}
            
            Fig.~\ref{fig:TdustAttenMagritte2D} shows the difference between the dust temperature, resulting from the attenuation approximation (see Sect.~\ref{sec:attenuation}; Fig.~\ref{Fig:2DplotAtten}, upper right panel), and the one obtained with \textsc{Magritte}, for a slice through the orbital plane. Close to the AGB star, the difference in dust temperature remains small, but the differences increase rapidly further out.
            In the shadow region behind the companion ($x>5$~au and $y<0$~au), the temperature in the attenuation approximation is too cold, but this scheme is able to reproduce the shadow, but the effect is exacerbated.
            
            The non-radial fraction of the radiation force, shown in Fig.~\ref{fig:radcompAtten}, is globally lower than in the case of the Lucy approximation, mainly because no accretion disk is forming in the attenuation simulation, and thus there is no region where the photons are trapped. When normalizing the radiation force to its maximum value (see Fig.\ref{fig:radcompscaledAtten}), we see a small non-radial contribution, similar to the Lucy case This renders the conclusion that the assumption of a radial radiation force is reasonably good.
            \begin{figure}
                \centering
                \includegraphics[width = \linewidth]{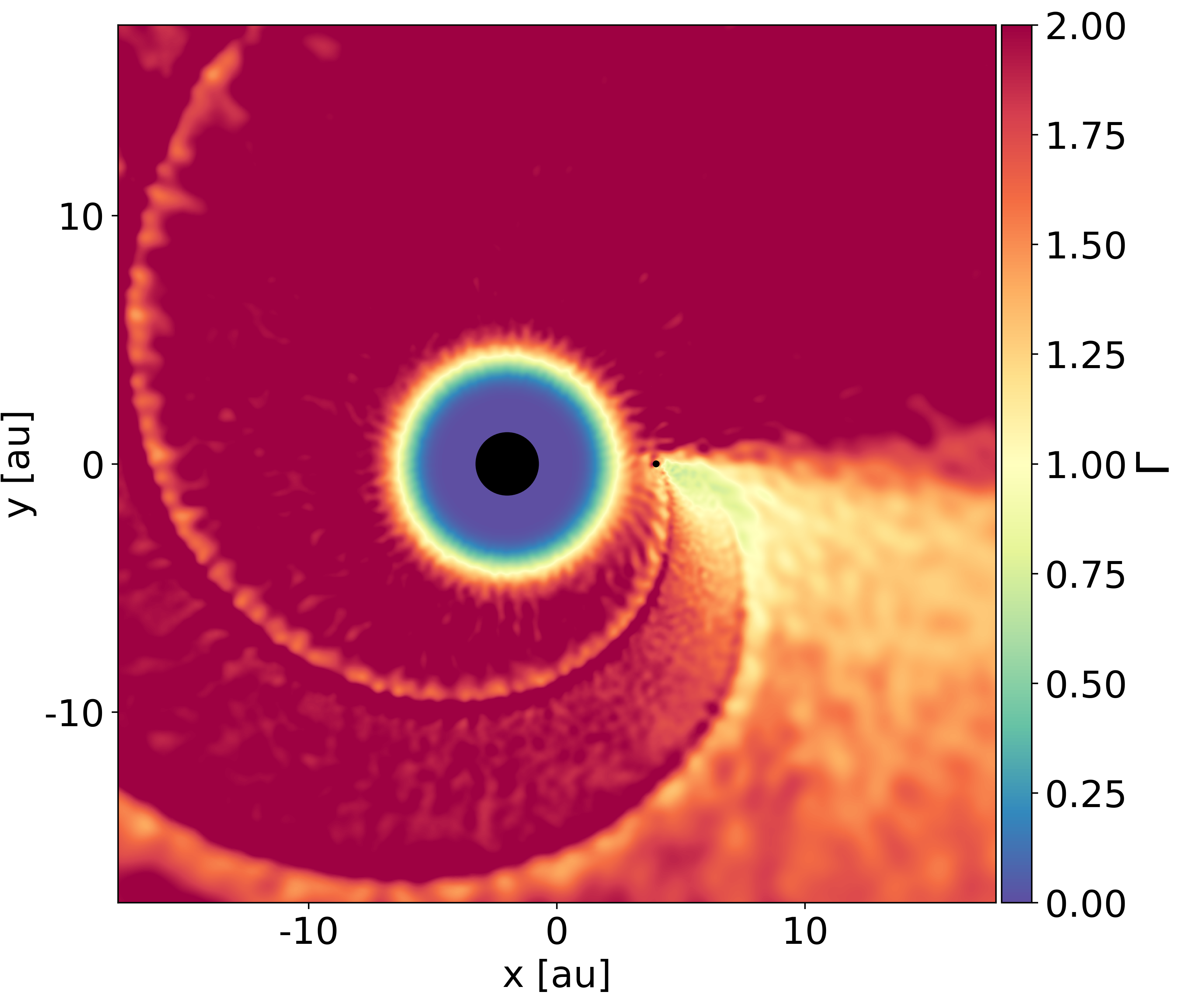}
                \caption{Eddington factor $\Gamma$ (Eq.~\ref{eq:Gamma}), calculated using the magnitude of the flux vector, in a slice through the orbital plane, using \textsc{Magritte} for the snapshot of the simulation, using the attenuation prescription.}
                \label{fig:radforceAtten2D}
            \end{figure}
            
            The Eddington factor $\Gamma$ (Eq.~\ref{eq:Gamma}) in the orbital plane is shown in Fig.~\ref{fig:radforceAtten2D}.
            Due to the higher temperature estimate in \textsc{Magritte}, the dust condensation radius is shifted farther out than in the attenuation approximation (Fig.~\ref{Fig:2DplotAtten}), resulting in a lower radiation force close to the AGB star. Further out and away from the shadow, the Eddington factor remains approximately constant according to \textsc{Magritte}, while it is monotonically decreasing in the attenuation approximation.
            Thus, the global radiation force is not correctly described by the attenuation approximation, for the same reason as the dust temperature.
            The shadow behind the companion ($x>5$~au and $y<0$~au) is reproduced by the attenuation approximation.
            However, due to the weak radiation force in this region, it is unclear whether this feature is modelled accurately or overcompensated.
            
            In summary, for regions that (locally) resemble a spherically symmetric wind (i.e.\ away from the companion), the Lucy approximation provides a correct estimate for the dust temperature and radiation force, taking into account the radiation that is absorbed and re-emitted isotropically.
            This effect heats up the material compared to the geometrical approximation.
            Close to the companion, dust is artificially heated due to the spherical symmetric assumption underlying the Lucy approximation.
            The Lucy approximation fails to account for the shadow cast behind the companion. 
            On the other hand, in the attenuation approximation, the regions away from the companion are not modelled correctly.
            The material is artificially cooled, causing a reduction of the radiation force.
            Close to the companion, a shadow forms, which is captured by the attenuation approximation both in the dust temperature as well as in the radiation force.
            However, the drastic artificial decrease of the radiation force, and the fact that most of the morphological simulation is spherically symmetric, favours the Lucy approximation.
            This suggests that a combination of the Lucy and attenuation approximation might yield even better results, combining the strengths of both approximations.

    \subsection{Future work}
        One of the ingredients still missing in this study is a treatment of pulsations.
        In our approach, pulsations are neglected and material is launched at a preset velocity from the stellar surface.
        In future studies, pulsations will be modelled by simulating a radially oscillating inner boundary acting as a piston.
        This method was originally used in the 1D study by \cite{bowen1988}, but has recently been implemented in the SPH context by \cite{Aydi2022}.
        After including pulsations and adopting the Lucy approximation from this study, the full \cite{bowen1988} study can be replicated in 3D, and include a companion.

        Furthermore, the treatment of the dust opacity can still be improved.
        \cite{Siess2022} already implemented carbon dust formation in \textsc{Phantom}, which can be used in combination with this study.
        Using this formalism, the dust opacities can be made more consistent with the physical and chemical environments of stellar outflows, which will result in a more accurate description of the radiation force.
        
        An accurate treatment of both chemical processes and cooling are of crucial importance as well \citep[e.g.][]{Boulangier2019, VandeSande2019}.
        The chemistry can alter the polytropic index, as well as the mean molecular weight, which now are assumed to be constant.
        To account for the complex chemistry taking place in AGB outflows, without trading too much of the required computation time, machine learning techniques will be used to emulate the chemical network \citep[e.g.][]{Holdship2021, Grassi2022}, reducing the computation time to allow for an on-the-fly simulation of the chemistry (Maes et al.\ in prep).
        Furthermore, we could improve on the cooling prescriptions. Cooling has a significant impact on the energy equation, since insufficient cooling can prevent the formation of accretion disks around the companion (\citealp{Theuns1993}, Malfait et al.\ in prep.).
        Moreover, it can alter the transfer rates of both mass and angular momentum.

        A final improvement for the simulations itself is the gas-dust interaction.
        In this study position coupling is assumed, although in a dust-driven wind the gas is dragged by the faster moving dust \citep{Mattsson2021}, which might result in significant drift velocities.
        In some cases this drag can lead to a decrease in both the mass-loss rate as well as the wind velocity \citep{Sandin2020}, which in turn influences the morphology as well.
        In \textsc{Phantom},  two approaches are already implemented to account for this drag.
        In the first approach, the two fluids are treated separately \citep{Laibe2012}, while in the second, both fluids can be combined using the so-called one-fluid approximation \citep{Laibe2014b,Laibe2014a, Price2015}.

        Once truly realistic models are acquired, these forward models can be used to compare to observations. This can be done using radiative transfer codes like \textsc{MCFOST} \citep{pinte2006,Tessore2021} or \textsc{Magritte} \citep{DeCeuster2020b, DeCeuster2020a} to produce synthetic observations. These synthetic observations are constructed by tracing chemical species in the simulation, and creating (synthetic) spectral line maps.
        These synthetic observations can then be compared to real observations, unraveling the origin of the complexities of these winds.

\section{Summary}\label{sec:summary}
    In this paper, we present the implementation of a ray tracer for radiative transfer in the SPH code \textsc{Phantom}.
    Using this ray tracer, a 3D map of the optical depth can be obtained, a key quantity that is needed, for example, in AGB wind simulations.
    This technique enables us to investigate four different radiative transfer approximations: the free-wind approximation (no radiative transfer), the geometrical approximation (including geometrical dilution, as well as dust formation), the Lucy approximation (including radiation not exclusively coming from the AGB star), and the attenuation approximation (including the absorption of the stellar flux without re-emission).
    
    We performed simulations of a single AGB star for each of the four prescriptions, with a low and high mass-loss rate, resulting in eight different models.
    The velocity profile, resulting from the free-wind approximation, remains relatively constant throughout the entire spatial domain.
    In the other approximations, where the radiation force and dust formation are accounted for, there is an initial decrease in velocity close to the AGB star, since this region is too hot for dust to condensate.
    Beyond the dust condensation radius, the material is accelerated outwards, allowing the velocity to escape the escape value.
    In the low mass-loss rate regime, the effects of the Lucy and attenuation approximations are small and the velocity profiles are almost identical.
    In the high mass-loss regime, the effects and their differences become more pronounced.
    For the Lucy approximation, the dust condensation radius moves outwards, resulting in a lower terminal velocity compared to the geometrical approximation.
    For the attenuation approximation, the dust condensation radius is shifted inwards, activating the radiation force earlier in the wind, but as radiation is attenuated, the radiation force also decreases, resulting in a flat velocity profile.
    
    We also considered binary systems and investigated the effect of the different radiative transfer approximations on the wind morphology.
    In the free-wind approximation, a thin, two-edged spiral structure forms, while in  the geometrical approximation, a thicker spiral is present, with increased interaction of the wind close to the companion.
    Again, in the low mass-loss rate regime the Lucy and attenuation approximation follow the geometrical approximation, while for the high mass-loss rate regime differences appear.
    In the Lucy approximation, an accretion disk is able to form around the companion because of the lower wind velocity.
    This accretion disk creates an additional bow shock, increasing the morphological complexity.
    In the attenuation approximation, the two-edged spiral is compressed significantly, such that the spiral arms interact, forming a single Archimedean spiral structure.
    
    In order to gauge the accuracy of these radiative transfer approximations, and hence the applicability, we compared the dust temperature and the radiation force from the simulations, using the Lucy and attenuation approximation, to results obtained with the full 3D radiative transfer code \textsc{Magritte}.
    We showed that the non-radial component of the radiation force is small, which implies that the often made assumption of a radial radiation force is adequate.
    The Lucy approximation can reproduce the dust temperature and the radiation field accurately in the parts of the simulation that resemble a spherically symmetric outflow.
    However, in the direction of the companion, the density is higher and the radiation field is overestimated due to the underlying assumption of spherical symmetry in the Lucy approximation.
    The Lucy approximation can model the radiation force correctly in regions that resemble a spherically symmetric outflow, but this approximation does not reproduce the shadow cast behind the companion because of excessive re-emission.
    In the attenuation approximation, the dust temperature and the radiation force are underestimated, because this approximation only models the extinction of radiation, but not the re-emission.
    Although, as expected, the attenuation approximation does account for the shadow cast behind the companion.
    In conclusion, the Lucy approximation turns out to be the most adequate radiative transfer prescription for AGB binary simulations, since most of the domain in the considered simulations resembles a spherically symmetric outflow, and the acceleration close to the AGB star is important for the companion interaction.

\begin{acknowledgements}
M.E., J.M., S.M. and L.D. acknowledge support from the Research Foundation Flanders (FWO) grant G099720N.
L.S. is a senior FRS-F.N.R.S research associate.
F.D.C is a Postdoctoral Research Fellow of the Research Foundation - Flanders (FWO), grant 1253223N.
T.K. and L.D. acknowledge support from the KU Leuven IDN grant IDN/19/028.
T.C. is a PhD Fellow of the Research Foundation - Flanders (FWO), grant 1166722N.
\end{acknowledgements}

\bibliographystyle{aa}
\bibliography{bibliography.bib}

\begin{appendix}
\section{Ray tracer}\label{app:raytracer}
    In order to investigate and validate the trade-offs made during the development of the ray-tracer (see Sect.~\ref{sec:NumImp}), we compare its performance against a second ray tracer algorithm that is optimized for accuracy, but not for computational speed.
    This ray tracer would be too slow to be used on-the-fly in \textsc{Phantom} simulations, but is useful to gauge the performance of our implementation.
    In this appendix, we gradually build up the reasoning that led to the ray tracer that was eventually implemented in \textsc{Phantom}, starting from this ideal (but slow) ray tracer (IRS; ideal ray-tracing scheme). 
    
    The ideal ray tracer works as follows.
    Starting from a specific location in the simulation, it calculates the nearest neighbours, out of which it will select the next particle to be considered on the ray.
    There, it recalculates the nearest neighbours, to again decide which particle should be considered next, and, as such, moves along the ray.
    To minimize the step size, and obtain the optimal discretization for integrals along the ray, the smallest set of neighbours is used.
    This set is obtained using a Delaunay tetrahedralization, which is used in \textsc{Magritte} as well \citep{DeCeuster2020a}.
    While walking through the selected particles along a ray, these particles are projected onto the ray, and, at these projected points, the integrand is evaluated (using Eq.~\ref{eq:integrand}).
    To calculate the optical depth, $\tau_\mathrm{IRS}$, in this ideal scenario, a ray is traced starting from the particle in question, in the direction of the AGB star.
    This ensures that each of the rays properly represents the actual integral that is calculated.
    The resulting algorithm is similar to \cite{Kessel2000}.

    To speed up the calculation, while trying to remain as accurate as possible, different improvements are investigated in the rest of this Section\footnote{All ray-tracing algorithms used in this appendix, can be found in the `utils' routines of \textsc{Phantom} (\url{https://github.com/danieljprice/phantom/blob/master/src/utils/utils_raytracer_all.F90}).}.
    To quantify the performance, the optical depth, $\tau$, calculated with our improved scheme, is compared to its IRS value, $\tau_{\text{IRS}}$, by calculating a relative error
    \begin{equation}\label{eq:relerror}
        \delta_{\rm rel} = \frac{1}{N_{\rm part}} \sum_{i=1}^{N_{\rm part}} \left|\frac{\tau_{\rm IRS, i} - \tau_i}{(\tau_{\rm IRS, i} + \tau_i)/2}\right|\ .
    \end{equation}
    To test the algorithm, we used an SPH model of an AGB binary, more specifically, dump 600 of model v05e50 ($v_{\rm inj} = 5\,{\rm km}\,{\rm s}^{-1}$, $e = 0.5$ using the free-wind approximation) of \cite{Malfait2021}. This is a late snapshot of the most complex morphology in that study, using $\approx 1.2 \times 10^6$ SPH particles. 
    
    \subsection{Nearest neighbours}\label{app:neighbours}
        A first bottleneck in the IRS algorithm, is the computation of the Delaunay nearest neighbours.
        Since \textsc{Phantom} already works with a $k$d-tree to store its particles (a tree-like data structure that significantly speeds up nearest neighbour determinations), calculating nearest neighbours leveraging this tree is much faster.
        The set of neighbours includes $\sim 60$  particles, such that the sampling along the ray can take larger steps, and hence the integration is done more crudely.
        Using this nearest neighbour set and comparing the results to the IRS, using Eq.~\ref{eq:relerror} results in a relative error of only 1\%, while giving a speed-up of a factor 2.
        This speedup does not yet include the calculation of the Delaunay nearest neighbours (taking a significant amount of time), such that the actual speed-up is even higher.
        
    \subsection{Ray directions}\label{app:raydirection}
        Tracing rays inwards is inefficient, since plenty of SPH particles will be passed several times, especially those close to the star.
        To avoid this, one can reverse the ray tracing by starting at the stellar surface, and pointing the rays outwards until the edge of the simulation is reached, instead of tracing a ray from each particle to the star.
        The rays that originate from the star, should be traced uniformly outwards.
        The directions of these rays can be determined using \textsc{HEALPix}\footnote{\url{https://healpix.sourceforge.io}} \citep{Gorski2005}.
        \textsc{HEALPix} subdivides the 2-sphere into iso-laterally distributed equal area pixels, such that the centre of each pixel can be used as the direction of a ray.
        The scheme starts with 12 pixels, so-called \textsc{HEALPix}-order 0, and the amount of rays can be increased by subdividing each cell into four, increasing the order $o$ by 1.
        This results in $n_{\rm rays} = 12 \times 4^{o}$.
        Once the rays are traced, the information along the rays needs to be mapped back to the SPH particles.
        Using \textsc{HEALPix}, the closest ray to a particle can be found easily by leveraging its smart pixel positioning.
        Once the nearest ray is found, the particle is projected on the ray, and the integral is evaluated there by linear interpolation between the two closest known values along the ray.

        \begin{figure}
            \centering
            \includegraphics[width=\linewidth]{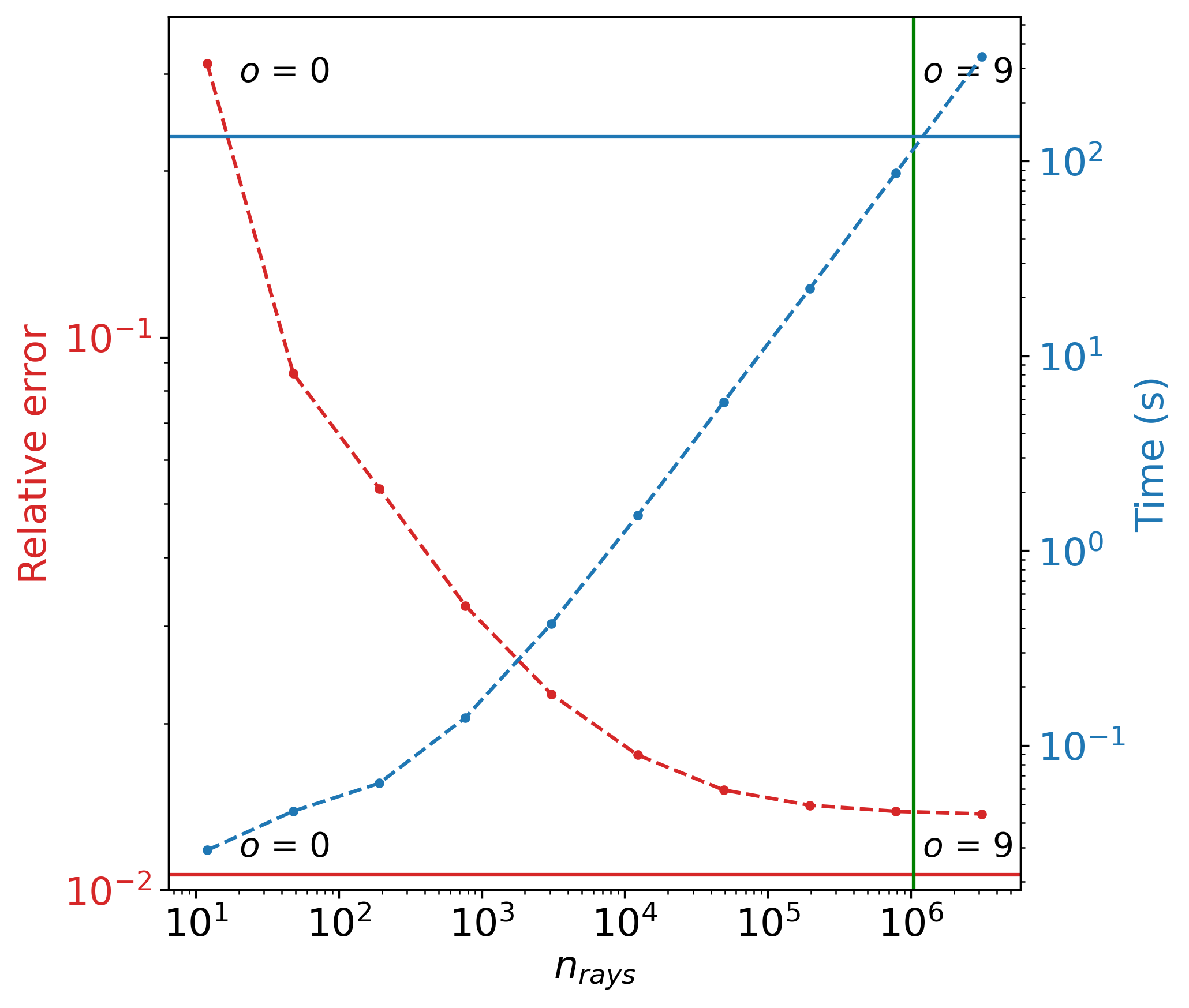}
            \caption{Relative error (Eq. \ref{eq:relerror}) and computation time, required for calculating $\tau$ as a function of the number of rays traced in the reference model. The solid red line represents the relative error using the SPH nearest neighbours, blue the computation time, and green the number of rays in the IRS case.}
            \label{fig:accuracydirection}
        \end{figure}

        The performance of this algorithm strongly depends on the \textsc{HEALPix} order. The relative error and the required computation time are shown as a function of the number of rays in Fig.~\ref{fig:accuracydirection}.
        There is an improvement in computation time, as long as the number of rays is less than the number of particles in the simulation.
        This is, however, at the expense of the relative error.
        For \textsc{HEALPix} order 5, for instance, the computation time goes down with a factor 250, while inducing a relative error of 2\%.
    
    \subsection{Ray interpolation}\label{app:rayinterpolation}
        To further reduce the 2\% relative error of the algorithm, we investigate the interpolation from the rays onto the particles.
        Instead of using only the information of the ray nearest to a particle, an interpolation between multiple rays can be used.
        We consider either the four or nine closest rays, as well as different interpolation exponents, $k$, in the interpolation formula, given by
        \begin{equation}
            I_i = \frac{1}{\sum_{j=1}r_j^{-k}}\sum_{j}\frac{I_j}{r_j^{k}}\ ,
        \end{equation}
        where $I_i$ is integral at the location of the $i$'th particle, $I_j$ is the integral at the projection of the $i$'th particle on the $j$'th ray, and $r_j$ the distance between these two. Here, $j$ is a sum over the four or nine closest rays, and $k$ specifies the interpolation exponent. 
        \begin{figure}
            \centering
            \includegraphics[width=\linewidth]{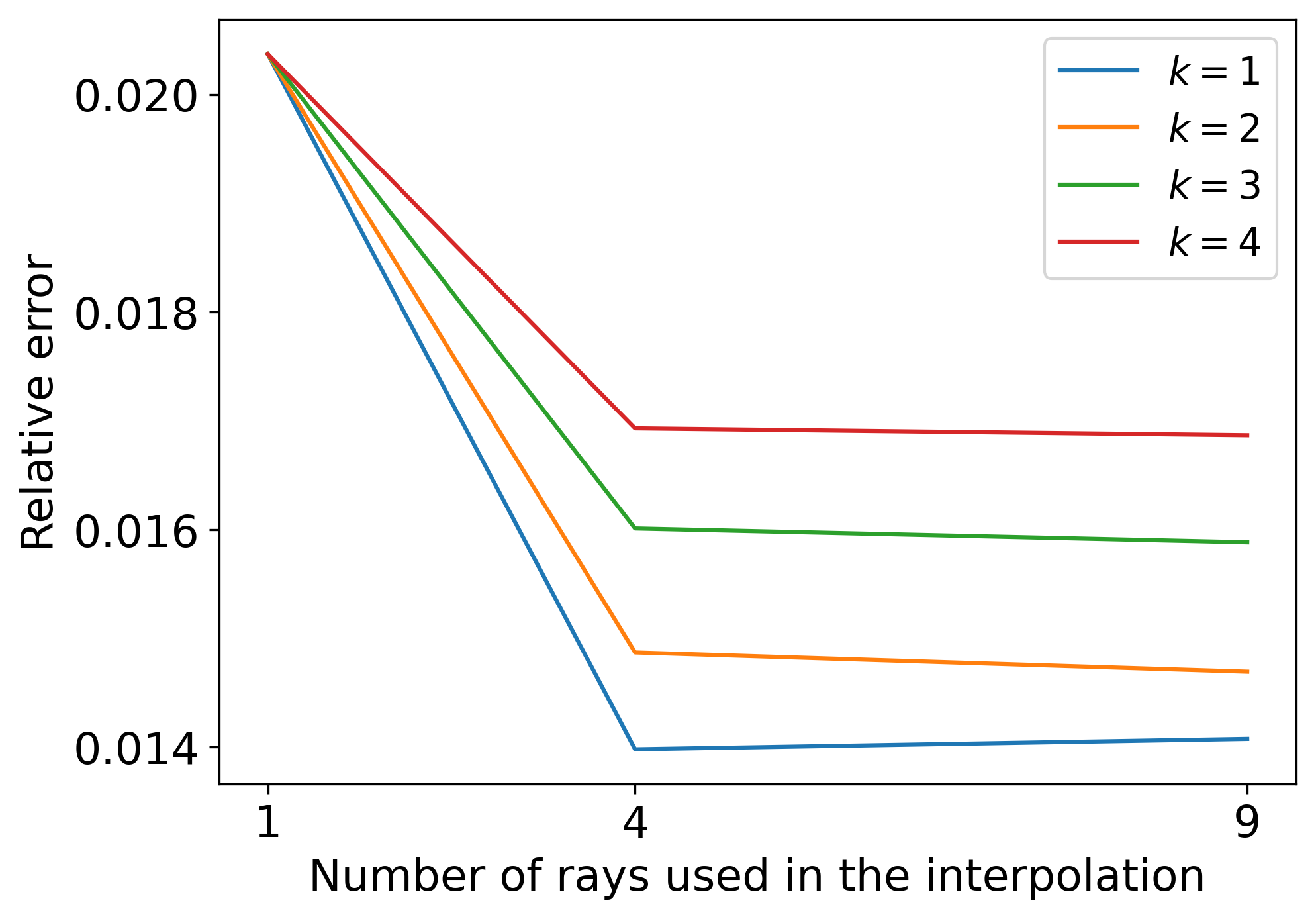}
            \caption{Relative error (Eq. \ref{eq:relerror}) as a function of the number of rays used in the interpolation (using \textsc{HEALPix} order 5).}
            \label{fig:interpolation}
        \end{figure}
        The relative error as a function of the number of rays, used in the interpolation for \textsc{HEALPix} order 5, is shown in Fig.~\ref{fig:interpolation}. There is a significant improvement when moving from one to four rays, as the latter interpolation can account for differences in between rays.
        However, going up to nine rays does not significantly improve the accuracy, as most of the smoothing already happens in between the four closest rays. The relative error even increases slightly when using $k = 1$, because the interpolation smooths out large variations.
        Since in our simulations large variation in density, and thus optical depth, are to be expected, using nine rays and $k=1$ will not be ideal.
        It turns out that for a interpolations exponent of $k=2$ and four rays is the best combination for the specific model that we considered.
        \begin{figure}
            \centering
            \includegraphics[width=0.85\linewidth]{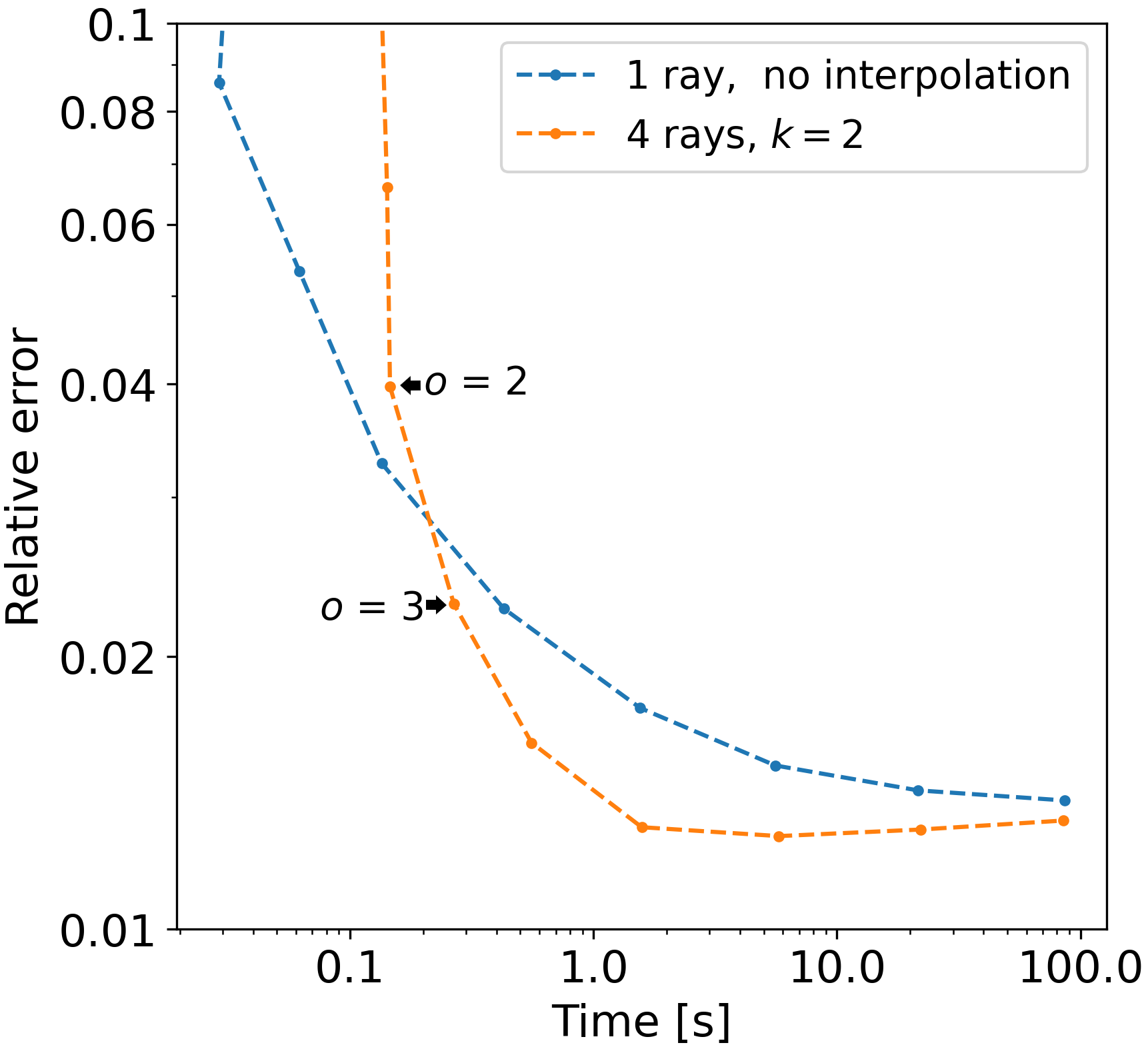}
            \caption{Relative error (Eq. \ref{eq:relerror}) as a function of computation time for calculating $\tau$. Different dots represent different \textsc{HEALPix} orders, where the blue represents no interpolation, and the orange represents the interpolation using four rays and $k=2$.}
            \label{fig:accuracyinterpolation}
        \end{figure}
        
        The resulting relative error as a function of computation time for this interpolation is shown in Fig.~\ref{fig:accuracyinterpolation}.
        For low \textsc{HEALPix} orders, the interpolation takes a significant amount of time, while only slightly improving the relative error.
        This is because the initial calculations along the few rays, that are traced, cannot capture all the complexities in between these rays.
        Going to higher orders (starting from \textsc{HEALPix} order 3) the interpolation is worthwhile.
        The increase in computation time becomes ever smaller, as the extra time used for the interpolation only scales with the number of particles, and not with the number of rays.
        Using \textsc{HEALPix} order 5, there is a speedup of a factor 250 compared to the IRS, while obtaining a relative error of only 1.5\%.

        Due to this significant speed-up, the calculation of the integrals using \textsc{HEALPix} order 5 only take about 10\% of the computation time of a normal SPH hydro timestep (or shorter for lower orders).
        Hence, the extra computation time to include these calculations is practically feasible to perform as on-the-fly calculations.

\section{Additional figures}
\begin{figure}[H]
    \centering
    \includegraphics[width=0.9\linewidth]{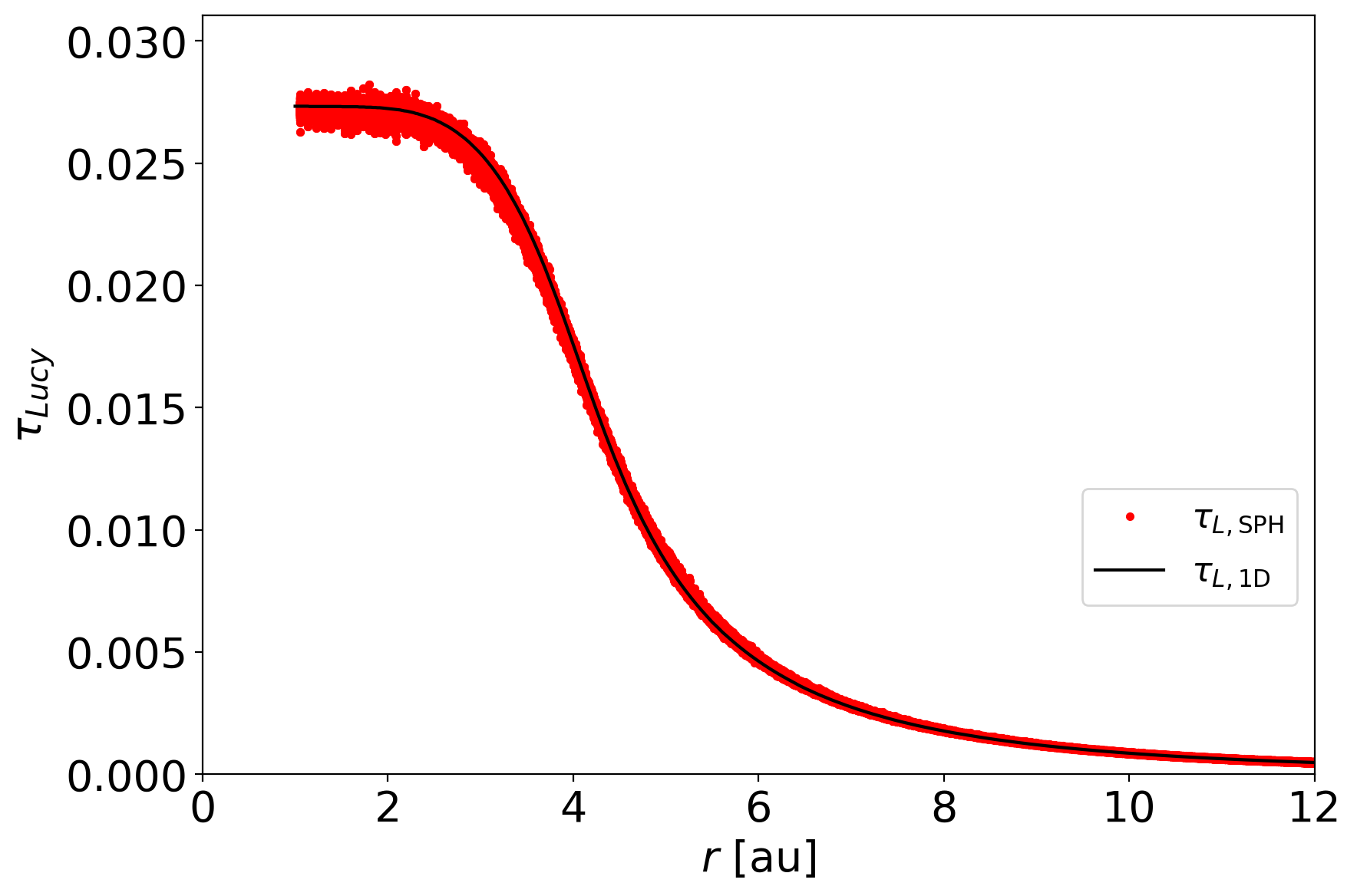}
    \caption{Lucy optical depth, $\tau_L$ (Eq.~\ref{eq:tauL}), as a function of radial distance in the single-star simulation, for the high mass-loss rate ($3 \times 10^{-6}\, {\rm M}_\odot \, {\rm yr}^{-1}$) model.}
    \label{fig:1DProfile_tau_lucy}
\end{figure}
\begin{figure}[H]
    \centering
    \includegraphics[width=0.9\linewidth]{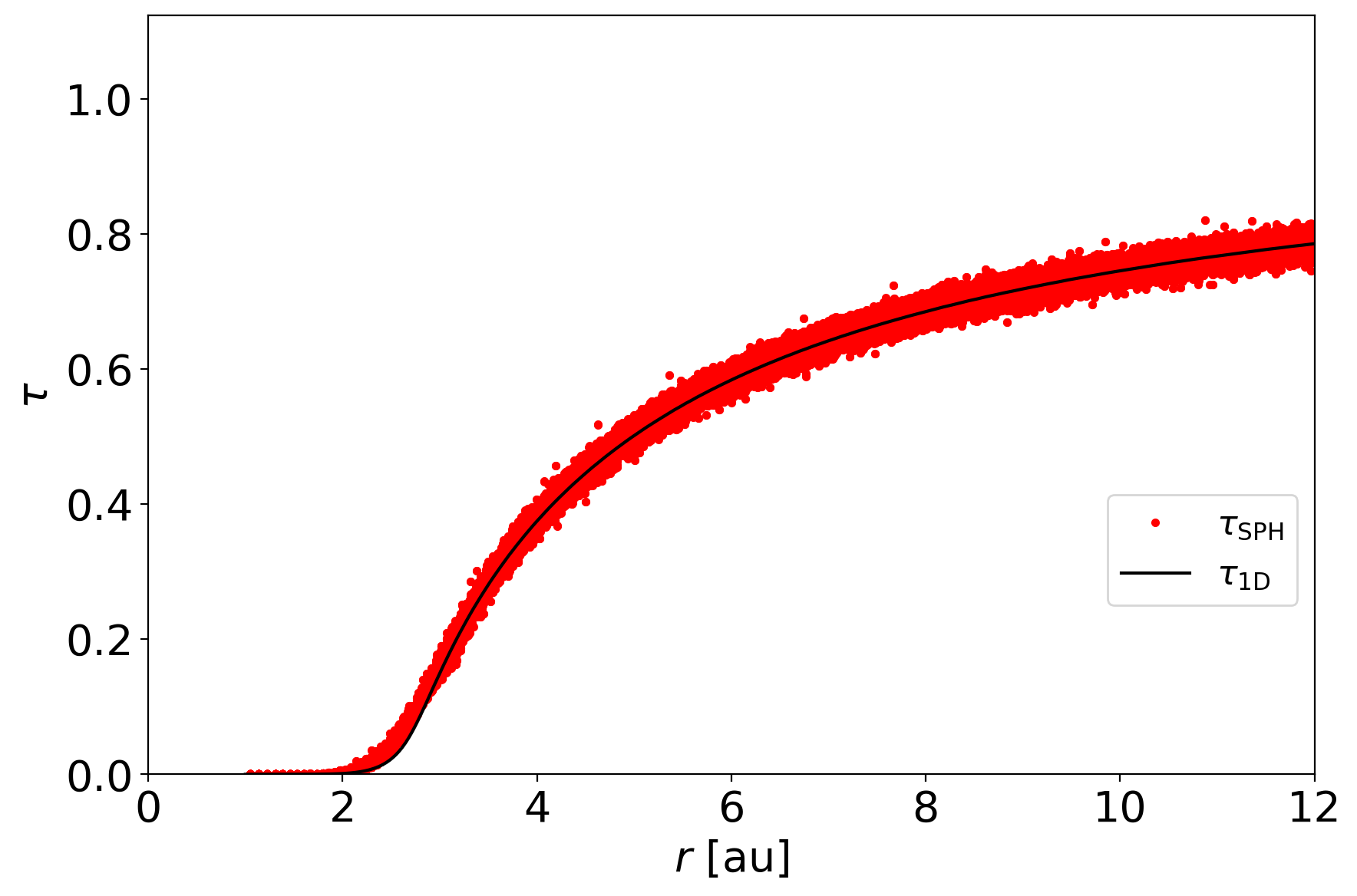}
    \caption{Optical depth, $\tau$ (Eq.~\ref{eq:tau}), in the attenuation prescription as a function of radial distance, for the high mass-loss rate  ($3 \times 10^{-6}\, {\rm M}_\odot \, {\rm yr}^{-1}$) model.}
    \label{fig:1DProfile_tau}
\end{figure}
\begin{figure*}
    \centering
    \includegraphics[width=.87\linewidth]{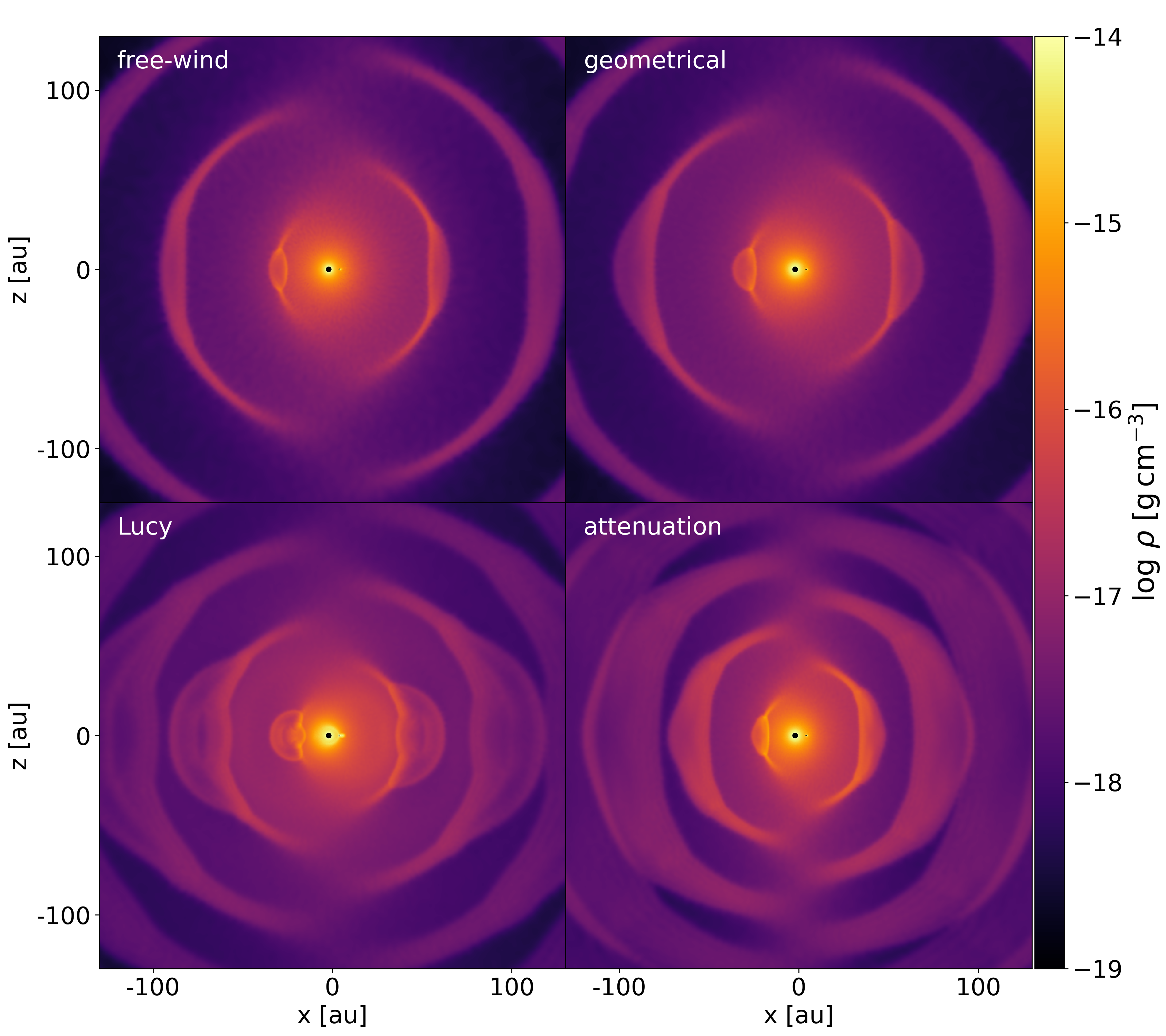}
    \caption{Density distribution in a slice through the meridional plane for the four prescriptions: free-wind (top left), geometrical (top right), Lucy (bottom left), and attenuation (bottom right) prescription, for the high mass-loss rate case with a binary companion. Both stars are on the x-axis, where the primary AGB star is on the left, and the companion on the right.}
    \label{Fig:2DdensityMeridional}
\end{figure*}
\begin{figure}
    \centering
    \includegraphics[width=0.9\linewidth]{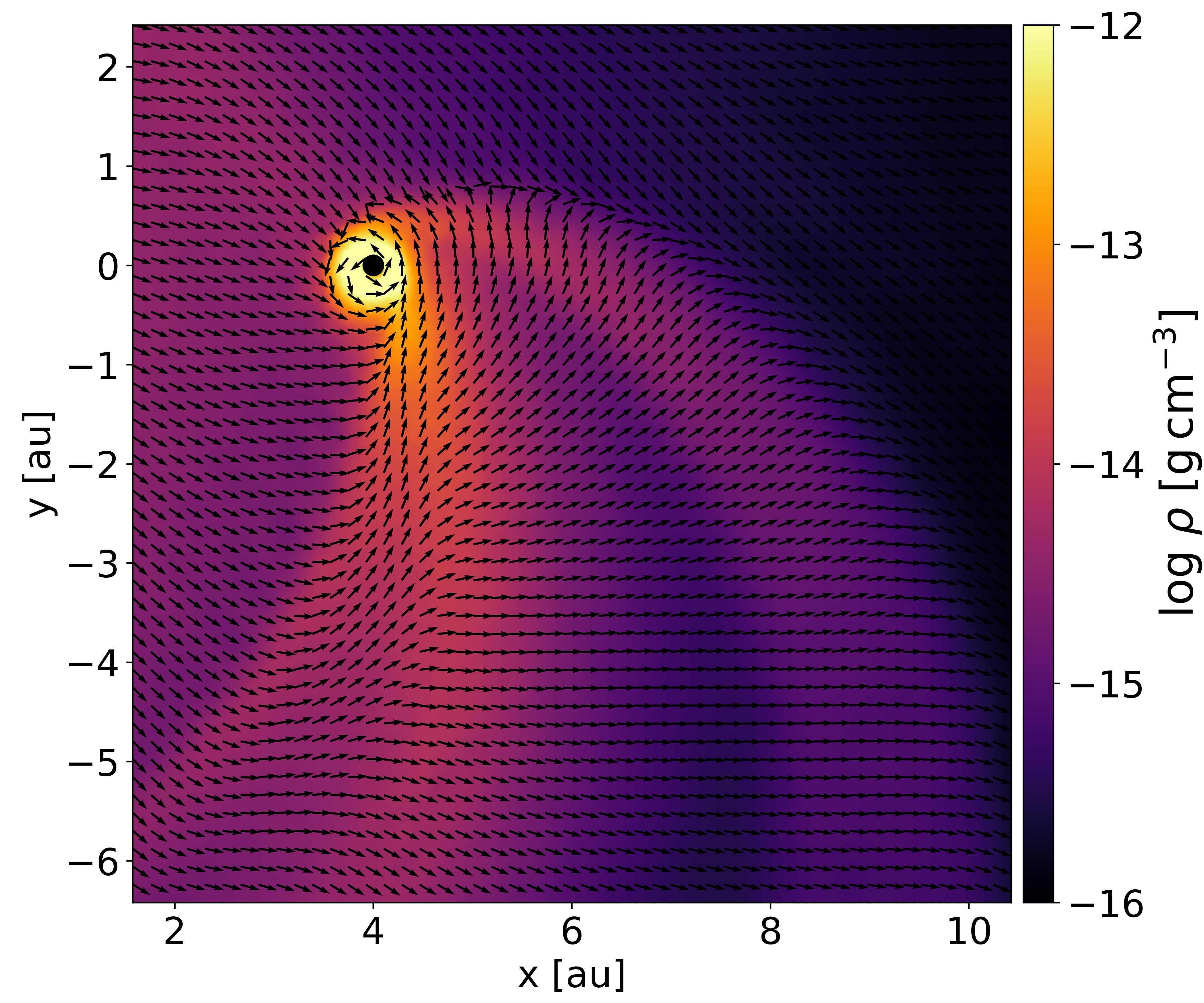}
    \caption{Density distribution in a slice through the orbital plane close to the companion in the Lucy approximation, revealing the presence of an accretion disk. The small arrows indicate the direction of the velocity field.}
    \label{fig:AccrDisk}
\end{figure}
\end{appendix}

\end{document}